%% file: PAPER_formal_JHEP_final.tex
\documentclass[12pt]{article}
\newcommand{\FlipTR}{UCR-TR-2023-FLIP-IG-11, CETUP-2023-004} 
\input{FlipPreamble}

\input{FlipAdditionalHeader} 
\input{FlipPreambleEnd}
\renewcommand{\k}{Q}

\begin{document}
\thispagestyle{firststyle} 	

\vspace*{1.5cm}

\begin{center}

    {\huge \textbf{Holography of Broken U(1) Symmetry } \par}
    
    \vskip 1.5cm

   { \bf
    Ian Chaffey$^{a}$,
    Sylvain~Fichet$^{b,c}$,
    and
    Philip Tanedo$^{a}$
    }
   \\
   \vspace{-.2em}
   { \tt \footnotesize
    \email{ichaf001@ucr.edu},
    \email{sfichet@caltech.edu},
  	\email{flip.tanedo@ucr.edu}
   }

   \begin{institutions}[1.7cm]
        \footnotesize
        $^{a}$
        {\textit{Department of Physics \& Astronomy,
         University of California, Riverside,
         \\ \phantom{$^{a}$}
         900 University Ave., Riverside,} 
         {\scalefont{.9}CA} \textit{92521}, {\scalefont{.9}USA}
         }
        \\
        \vspace*{0.05cm}
        $^{b}$
        {{\scalefont{.9}ICTP} \textit{South American Institute for Fundamental Research} \textit{\&}
        {\scalefont{.9}IFT-UNESP}
        \\ \phantom{$^{b}$}
        \textit{R.~Dr.~Bento Teobaldo Ferraz 271, S\~ao Paulo, Brazil}
        }
        \\
        \vspace*{0.05cm}
        $^{c}$
        {\textit{Centro de Ciencias Naturais e Humanas,
        Universidade Federal do}
         {\scalefont{.9}ABC},
         \\ \phantom{$^{c}$}
         \textit{Santo Andre,}  \textit{09210-580 S\~ao Paulo, Brazil}
        }
    \end{institutions}

\end{center}

\vskip 1.5cm

\begin{abstract}
\noindent 

We examine the Abelian Higgs model in $(d+1)$-dimensional anti-de Sitter space with an ultraviolet brane. 
The gauge symmetry is broken by a bulk Higgs vacuum expectation value triggered on the brane.
%
We propose two separate Goldstone boson equivalence theorems for the boundary and bulk degrees of freedom.
%
We compute the holographic self-energy of the gauge field and show that  its spectrum  is either a continuum, gapped continuum, or a discretuum as a function of the Higgs bulk mass.
%
When the Higgs has no bulk mass, the \AdS isometries are unbroken. We find in that case that the dual \CFT has a non-conserved \UU current whose anomalous dimension is proportional to the square of the Higgs vacuum expectation value.  
%
When the Higgs background weakly breaks the \AdS isometries, we present an adapted \WKB method to solve the gauge field equations. We show that the \UU current dimension runs logarithmically with the energy scale in accordance with a nearly-marginal \UU-breaking deformation of the \CFT.  



\end{abstract}

\small
\setcounter{tocdepth}{2}
\newpage
\tableofcontents
\normalsize


\newpage
\section{Introduction}
\addtocontents{toc}{\protect\setcounter{tocdepth}{1}}

The Abelian Higgs model is a cornerstone of quantum field theory. It is the canonical model that provides the key insights on the spontaneous symmetry breaking of a gauge symmetry and the behavior of massive gauge fields.  In this paper we study elementary aspects of the Abelian Higgs model in anti-de~Sitter spacetime (\AdS).  

Aspects of broken gauge symmetry in \AdS have been studied in the past, typically in a slice of \AdS and in the context of the Randall--Sundrum~1 model~\cite{Randall:1999ee}, or in the presence of a charged black hole  leading to holographic superconductor models~\cite{Hartnoll:2008kx}. However, there has been relatively little work on the field-theoretical aspects of the Abelian Higgs in \AdSdpp. This work fills that gap.

In our \AdS Abelian Higgs model, the bulk of \AdSdpp contains a \UU gauge field and a charged scalar field, the Higgs field.  We assume that there exists a $(d-1)$-dimensional brane towards the \AdSdpp boundary, the \UV brane.  A potential localized on this brane induces a nonzero Higgs vacuum expectation value (\vev) that extends into the bulk. 

One motivation for this study is to understand the curved space counterpart of a set of well-known phenomena in flat space. For instance, one would like to see how the Goldstone boson equivalence theorem manifests itself in \AdSdpp.  Moreover, in \AdS the Higgs background value can break the isometries of the bulk spacetime. One would like to see the impact of this nontrivial background on the gauge field.

Another motivation is the gauge--gravity correspondence. Placing probes on the \UV brane defines a holographic view of the \AdS Abelian Higgs model. When the \AdSCFT correspondence applies, one would like to understand the properties of the holographic \CFT dual, such as the fate of the dual global $\UU$ current.

We briefly review past works involving gauge fields in \AdS.

\paragraph{Holography and dual conformal theory.}
    
From the viewpoint of sources placed on the \AdS boundary or  the \UV brane ({i.e.} the regulated \AdS boundary), the bulk physics gives rise to a conformal theory  with a large number of color $N$ and large 't~Hooft coupling, see {e.g.} \cite{Aharony:1999ti,Zaffaroni:2000vh,Nastase:2007kj,Kap:lecture} for general \AdSCFT references. Aspects of unbroken gauge fields in \AdSCFT are discussed in {e.g.} \cite{Cacciapaglia:2008ns,Friedland:2009iy,Friedland:2009zg}. 
Here we analyze the holographic dual of a gauge field broken by an arbitrary bulk vacuum expectation value (\vev). There are relatively few \AdSCFT studies about field-theoretical aspects of broken internal symmetries, notably Ref.~\cite{Anand:2015zea} studies aspects of the Goldstone boson equivalence theorem. The Abelian Higgs model in the presence of charged black hole has been exploited to build models of holographic superconductors and superfluid, see e.g.~\cite{Gubser:2008px,Salvio:2012at,Horowitz:2013jaa,Hartnoll:2016apf,DeWolfe:2018dkl}. We study the zero temperature (no black hole) case and focus on the quantum fields living in the \AdS--Higgs background.

\paragraph{Warped extra dimensions and  hidden sectors}

Aspects of unbroken gauge fields in \AdSdpp spacetimes---or in a truncated versions of it---are studied in {e.g.}~\cite{Davoudiasl:1999tf, Pomarol:1999ad,Randall:2001gb,Randall:2001gc, Batell:2005wa, Batell:2006dp, 
Cacciapaglia:2008ns,Friedland:2009iy,Friedland:2009zg}. 
One may break gauge symmetry on a \FiD slice by imposing appropriate brane boundary conditions, see {e.g.}~\cite{Csaki:2003dt, Csaki:2005vy} and \cite{Cui:2009dv, McDonald:2010iq, McDonald:2010fe} for applications.
Refs.~\cite{Chang:1999nh, Huber:2000fh, Davoudiasl:2005uu, Cacciapaglia:2006mz,Archer:2012qa,Archer:2014jca} construct explicit models with a bulk Higgs in the Randall--Sundrum scenario; though electroweak naturalness typically leads to a focus on the case of a Higgs \vev that is either restricted or otherwise highly localized to an \IR brane.

From the viewpoint of an \UV brane, a warped spacetime can provide sectors that are naturally light, weakly coupled to the Standard Model, and have strong self interactions. Such a framework is a natural way to build strongly-interacting hidden sectors. This fact is emphasized in \cite{Brax:2019koq}, see also \cite{McDonald:2010fe,Gherghetta:2010cq,McDonald:2010iq,vonHarling:2012sz} for earlier related works. By the \AdSCFT correspondence, the setup amounts to having a \FoD (nearly) conformal hidden sector. 

\paragraph{Outline}

We define the \AdS Abelian Higgs model and derive the equations of motion and propagators in Sections~\ref{sec:model} and \ref{sec:propagators}.  
In subsequent sections, we apply these basic results to a series of semi-independent examinations of different aspects of the theory.
Section~\ref{sec:GET} examines the notion of a Goldstone equivalence theorem. Section~\ref{sec:landscape} investigates the gauge boson boundary action as a function of the Higgs \vev profile using either exact solutions or a \WKB approximation. In Section~\ref{sec:CFT_Comments}, we study the properties of the dual \CFT model with a broken \UU symmetry. Finally, in Section~\ref{eq:CFTapprox} we develop an adapted \WKB method to compute the gauge boson boundary action in a \vev configuration that approximately respects the \AdS isometries. Fig.~\ref{fig:summary} provides a summary of the main theoretical results.

\section{\texorpdfstring{A Broken U(1) in AdS$_{d+1}$}{A Broken U(1) In AdS}}
\label{sec:model}
\addtocontents{toc}{\protect\setcounter{tocdepth}{2}}

We consider a \UU gauge symmetry that is broken by the vacuum expectation value of a scalar field in anti-de~Sitter spacetime. The full action takes the form
\begin{align}
    S
    =
    \SD
    +
    S_{\textnormal{fix}}^{\textnormal{bulk}}
    +
    S_{\textnormal{fix}}^{\textnormal{UV}}\, .
    \label{eq:full_action}
\end{align}
$\SD$ includes the kinetic terms and scalar potential. $S_{\textnormal{fix}}^{\textnormal{bulk}}$ and $S_{\textnormal{fix}}^{\textnormal{UV}}$ fix the gauge.

\subsection{Geometry}

\begin{figure}
    \centering
    \includegraphics[height=0.4\textwidth]{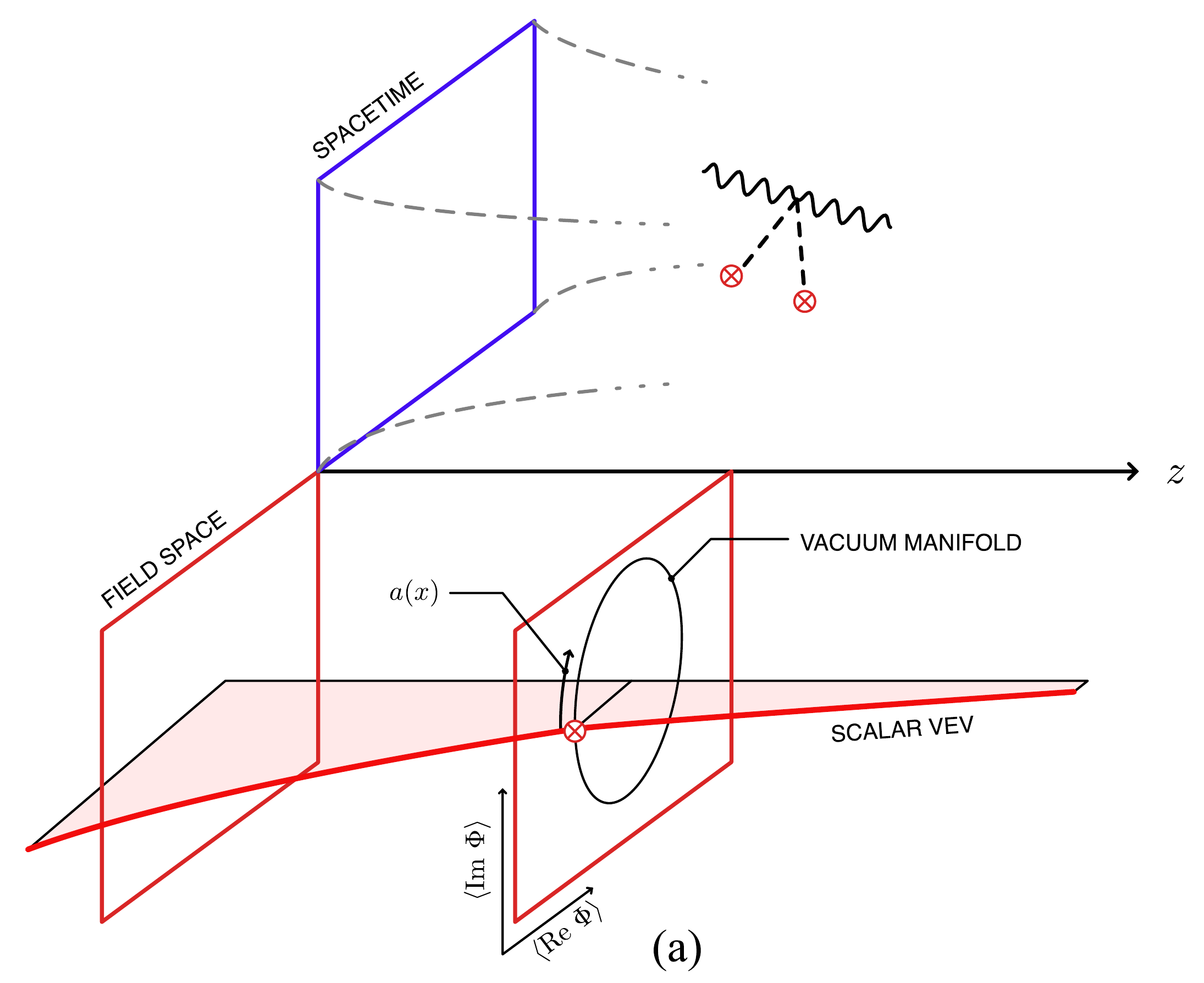} \quad
    \includegraphics[height=0.4\textwidth]{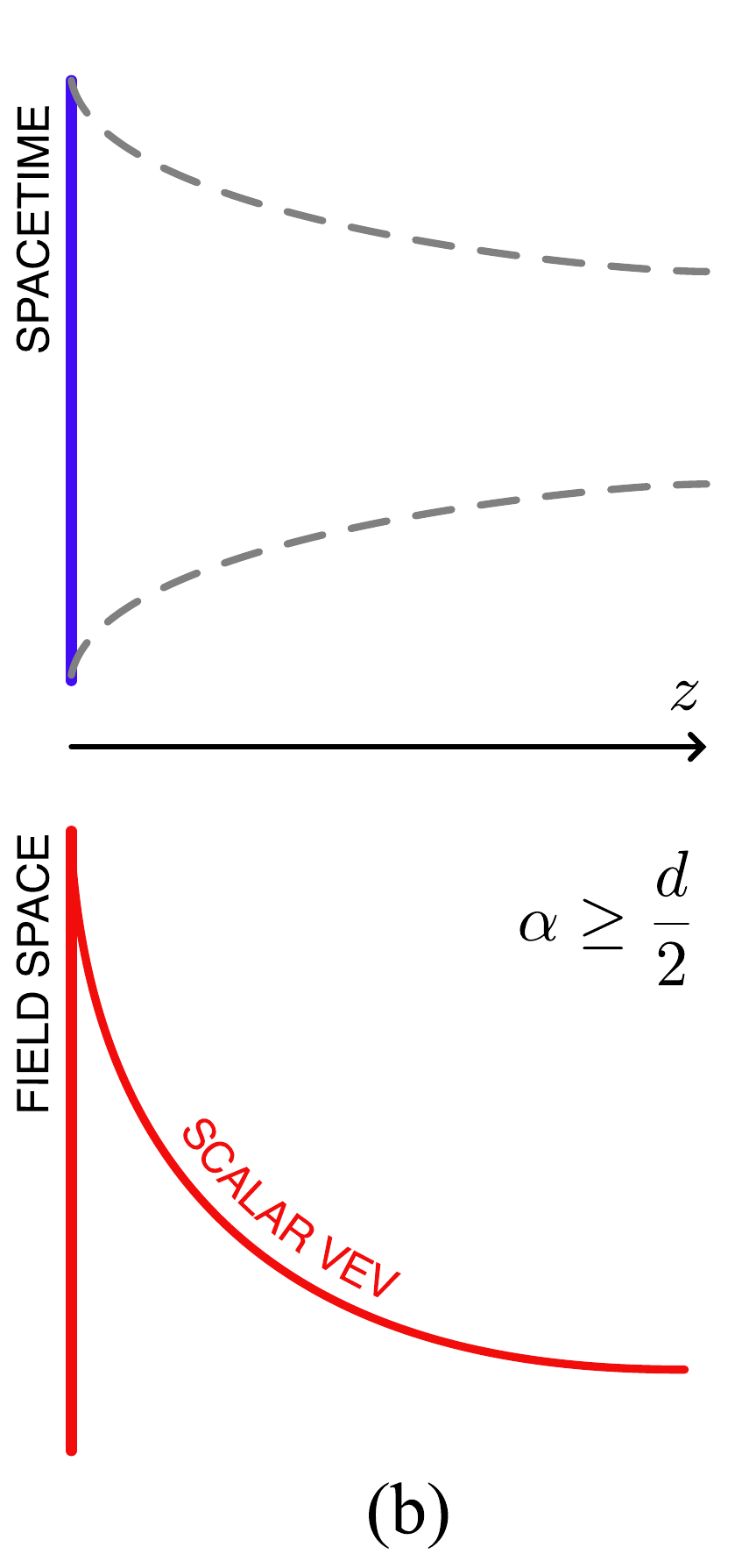} \;\;
    \includegraphics[height=0.4\textwidth]{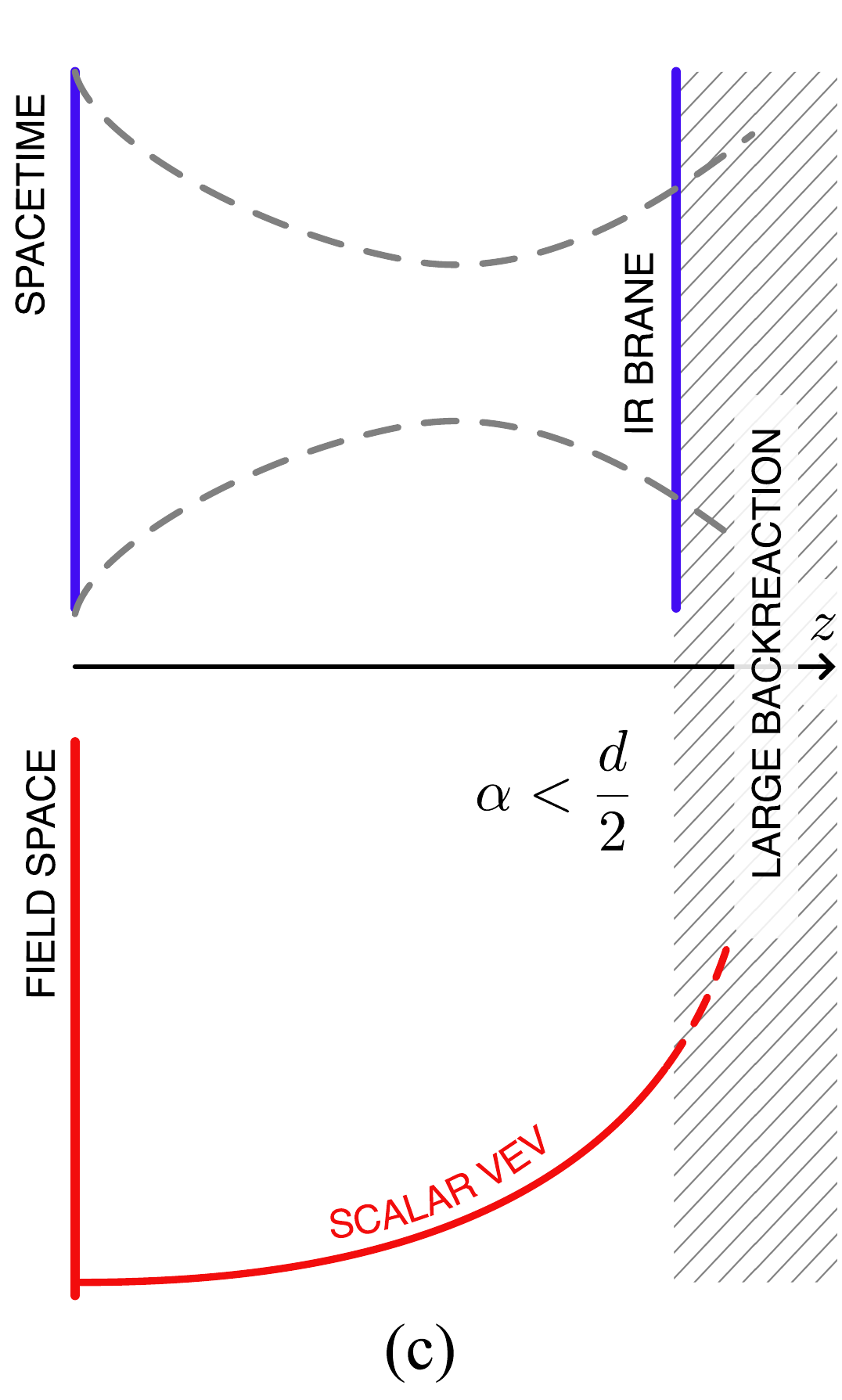}
    \caption{(a) Sketch of the model in spacetime (top) and field space (bottom, red slices).  The dashed gray lines depict the scaling of the metric. Red crosses represent \vev insertions. The Goldstone direction along the scalar vacuum manifold is shown for a particular value of $z$. For illustrative purposes we assume  $\alpha \geq d/2$.  (b) Same as (a), projected onto two dimensions. (c) Two dimensional projection for $\alpha > d/2$ where the scalar \vev induces a large backreaction. We model this with an infrared (\IR) brane.}
    \label{fig:setup}
\end{figure}

We work in the Poincar\'e patch of $(d+1)$-dimensional anti-de~Sitter spacetime, \AdSdpp. The metric in the conformal frame is
\begin{align}
    \D{s}^2
    = g_{MN}\D{x}^N\D{x}^N 
    = \left(\frac{R}{z}\right)^{\!2} \left(\eta_{\mu \nu}\D{x}^\mu \D{x}^\nu-\D{z}^2\right)
    \label{eq:metric} \, ,
\end{align}
where $R$ is the \AdS radius of curvature and $\eta_{\mu \nu}$ is the $d$-dimensional Minkowski metric, $\eta_{\mu\nu}=\textnormal{diag}(1,-1,\dots ,-1)$.  Lowercase Greek indices ($\mu, \nu, \cdots$) run from $0$ to $(d-1)$ and uppercase Roman indices ($M, N, \cdots$) run from $0$ to $d$. The $z$ coordinate is restricted to $z\geq \zUV$. 

We assume the existence of a codimension-one domain wall at $z=\zUV$ called the \UV~brane. We assume it is a boundary of spacetime, which can be understood as a regularization of the conformal boundary of \AdS to a well-behaved boundary.
We use the terms `\UV~brane' and `boundary' interchangeably, with a preference for \UV~brane when referring to our specific model and boundary for more general concepts.
We assume that the \UV brane is static and set its position to $\zUV=R$ without loss of generality.\,\footnote{Throughout the paper we only use $\zUV=R$ to simplify $\sqrt{|g_\textnormal{ind}|}=(R/\zUV)^d$.  We write $\zUV$ and $R$ separately elsewhere. Expressions with arbitrary $\zUV$ can be simply  obtained  by restoring  $\sqrt{|g_\textnormal{ind}|}$ in the calculations.} In these coordinates the bulk and \UV~brane integration measures are
\begin{align}
\sqrt{|g|}~\Ddppx
    &= \left(\frac{R}{z}\right)^{\!d+1}
    \Ddppx
    &
\sqrt{|g_\textnormal{ind}|}~\Ddx
    &= \operatorname{d}^{d}\!x \ . \label{eq:rootg}
\end{align}

The bulk scalar field has a vacuum expectation value (\vev) that breaks the \UU gauge symmetry. Depending on the shape of the \vev along the $z$-direction, the energy density of the \vev can deform the metric. A full solution to the $(d+1)$-dimensional gravity--scalar system depends on the specific choice of bulk potential and is beyond the scope of this paper. 
Instead, we simply model this backreaction with an \IR brane that truncates the bulk for large values of $z$ where the backreaction is large. We further assume that this \IR brane is stabilized, for example by the Goldberger--Wise mechanism~\cite{Goldberger:1999uk}. 
We sketch this model in Fig.~\ref{fig:setup} and provide parametric estimates for the backreaction and \IR brane location in Section~\ref{sec:Stability}.

For dimensions $d\geq 3$ there are non-renormalizable interactions with the bulk gravitons. In that case, the model is understood to be a low-energy effective field theory (\EFT). The cutoff scale of the \EFT is tied to the $(d+1)$-dimensional Planck scale $M_*$.
The \EFT is valid when the product of the \AdS curvature and $(d+1)$-dimensional Planck scale satisfies $RM_* \gtrsim 1$ and when the couplings are at most order one in units of $R$ times appropriate loop factors. We refer to Refs.~\cite{Chacko:1999hg,Agashe:2007zd,Costantino:2020msc} for more details.

\subsection{U(1) Action}

The $(d+1)$-dimensional action for a \UU gauge field $A_M=(A_\mu,A_z)$ and complex scalar $\Phi$ is 
\begin{align}
    \SD
    & =  
    \int \left(\frac{R}{z}\right)^{\!d+1} \! 
    \Ddppx
    \, 
        \mathcal{L}
        \,+
    \int_\textnormal{UV}
     \Ddx
     \,\mathcal{L}_{\textnormal{UV}}
    \ ,
    \label{eq:SD}
\end{align}
where we separate the bulk $\mathcal L$ and brane-localized $\mathcal{L}_{\textnormal{UV}}$ Lagrangians.  We write $\int_\textnormal{UV}$ for integration restricted to the \UV~brane. 
The bulk Lagrangian is
\begin{align}
    \mathcal{L}
    &=
    -\frac{1}{4{\g}^2}F_{M N}F^{MN}
   +
    \left( \smash{ D_M \Phi } \right)^{\dagger}\!
    \left( \smash{ D^M \Phi } \right)
    -
    \mu^2 \Phi^\dagger \Phi 
    &
    D_M &=\partial_M-i A_M
    \,.
    \label{eq:bulk:Lagrangian}
\end{align}
${\g}$ is the $(d+1)$-dimensional gauge coupling. The uppercase Roman indices are raised using the inverse metric $g^{MN}\!$. We choose a normalization where the mass dimensions of the gauge field, scalar, and gauge coupling are
\begin{align}
    \left[A_M\right] &= 1
    &
    \left[\Phi\right] &= \frac{d-1}{2}
    &
    \left[{\g}\right] &= \frac{3-d}{2} \ .
\end{align}
Perturbativity of the effective theory imposes ${\g}^2R^{3-d}\lesssim 1$.
The brane-localized Lagrangian is
\begin{align}
    \mathcal{L}_{\textnormal{UV}}
    &=
    -\frac{\auv R}{4{\g}^2}
    F_{\mu\nu}F^{\mu\nu}
    +
    c_{\textnormal{UV}} R 
    \left(D_\mu \Phi\right)^{\dagger}\!\left(D^\mu \Phi\right)
    -
    V_{\textnormal{UV}}[\Phi] \ ,
    \label{eq:LUV}
\end{align}
where the lowercase Greek indices are contracted with the $d$-dimensional Minkowski metric, $\eta^{\mu\nu}$. We pull out factors of $R$ to facilitate dimensional analysis. $\auv$ and $\cuv$ are dimensionless coefficients that control the extent to which the bulk fields are localized towards the brane. The limit where these coefficients go to infinity correspond to brane-localized fields.

\subsection{Spontaneous Symmetry Breaking}

The  {homogeneous} bulk equation of motion for $\Phi$ is
\begin{align}
    \left[
    -\partial^2+z^{d-1}\partial_z \left(\frac{1}{z^{d-1}}\partial_z\right)
    -\left(\frac{\mu R}{z}\right)^{\!2}
    \right] \Phi(x,z)
    =
    0 \ .
    \label{eq:Phi:EOM}
\end{align}
Here $\partial^2$ is the Minkowski Laplacian with respect to the $d$ dimensions transverse to $z$. We write $|_{\zUV}$ to indicate a quantity evaluated on the \UV brane, $z=\zUV$. The field satisfies the boundary condition
\begin{align}
\left.
    \partial_z \Phi
    -
    c_{\textnormal{UV}}  R\, \partial^2\Phi
     -
     V_\textnormal{UV}'[\Phi]\;
     \right|_{\zUV}
     &=
     0
     &
     V_\textnormal{UV}'[\Phi]&\equiv
     \frac{\partial V_\textnormal{UV}[\Phi]}{ \partial \Phi}
     \ ,  \label{eq:BC_vev}
\end{align}
We include a spontaneous symmetry-breaking potential only on the \UV brane, \eqref{eq:LUV}. The bulk equation of motion \eqref{eq:Phi:EOM} dictates how  the {symmetry breaking} extends into the bulk. The brane-localized scalar potential is\footnote{In this work we focus on the case of a brane-localized potential that generates a $\Phi$ vev that extends into the bulk.
It is, in principle, possible to use a bulk potential to generate the \vev. As long as the bulk potential is localized towards the brane, it effectively sets a boundary condition for the $\Phi$ vev and we expect the results to be qualitatively similar to this paper. Beyond this limit, a thorough analysis is required. The intuition of the present paper requires the geometry remain asymptotically AdS near the brane.}
\begin{align}
    V_{\textnormal{UV}}\left[\Phi\right]
    =
    -
    m_{\textnormal{UV}}^2 R |\Phi|^2
    +
    \lambda_{\textnormal{UV}} R^{d-2} |\Phi|^4
    \, . 
    \label{eq:V_UV}
\end{align}
This potential induces a finite \vev for $\Phi$ that spontaneously breaks the \UU symmetry,
 \begin{align}
     \langle\Phi(x,z)\rangle
     &=
     \frac{v(z)}{\sqrt{2}}
     &
     \Phi(x,z)
     &=
     \frac{h(x,z)+v(z)}{\sqrt{2}}
     \e^{i\frac{\pi(x,z)}{v(z)}}
     \ .
     \label{eq:Phi:vev:and:CCWZ}
 \end{align}
The warping of the space in the extra dimension induces a $z$-dependence on the \vev. 
On the right-hand side of \eqref{eq:Phi:vev:and:CCWZ}, we parameterize $\Phi$ nonlinearly with respect to a radial mode $h(x,z)$ and a Goldstone mode $\pi(x,z)$~\cite{Coleman:1969sm,Callan:1969sn}.  

The $\Phi$ equation of motion \eqref{eq:Phi:EOM} and boundary condition \eqref{eq:BC_vev} determine the \vev profile:
\begin{align}
    \left[
        z^{d-1} \partial_z 
        \left( \frac{1}{z^{d-1}} \partial_z \right)
        -
        \left( \frac{\mu R}{z} \right)^{\!2}
    \right] v(z) &= 0 
    &
    \left.
    \partial_z v(z)
    \right|_{\zUV}
    &=
    V_\textnormal{UV}'[v(\zUV)]
    \ .
    \label{eq:eom:vev}
\end{align}
The \vev has a general bulk solution
\begin{align}
    v(z)
    & =
    B
    \left(\frac{z}{R}\right)^{\tfrac{d}{2}-\alpha}
    +
    C \left(\frac{z}{R}\right)^{\tfrac{d}{2}+\alpha}
    &
    \alpha^2
    &=
    \frac{d^2}{4}+\mu^2 R^2
    \, .
    \label{eq:vev:boundary}
\end{align}
The bulk equation of motion does not specify the sign of $\alpha$. The two terms in \eqref{eq:vev:boundary} reflect this since they are related by $\alpha \leftrightarrow -\alpha$. Requiring the regularity of the \vev for $z\to\infty$ removes the solution that grows more quickly. This is equivalent to selecting $\alpha>0$ and setting $C=0$.
The boundary condition on the right-hand side of \eqref{eq:vev:boundary} fixes the \vev normalization:
\begin{align}
    v(z)
    & =
    \vuv
    \left(\frac{z}{R}\right)^{\tfrac{d}{2}-\alpha}
    &
    \vuv
    &=
    \left(\frac{R}{\zUV}\right)^{\tfrac{d}{2}-\alpha}
    \sqrt{
        \frac{ \frac{d}{2}-\alpha+m_{\textnormal{UV}}^2 R^2 }{ \lambda_{\textnormal{UV}}R^{d-1} }
        }
    \ .
     \label{eq:vevsol}
\end{align}
It is convenient to factor out the $(R/\zUV)^{\frac{d}{2}-\alpha}$ in the definition of $\vuv$. A nonzero \vev exists only if $\frac{d}{2}-\alpha+m_{\textnormal{UV}}^2 R^2 >0$. The appearance of this combination of parameters is equivalently understood from the holographic potential that encodes the dynamics of the symmetry breaking on the brane, see Section~\ref{sec:effective:UV:potential}.

\subsection{Gravitational Backreaction and an IR Brane}
\label{sec:Stability}

The scalar \vev is a classical background whose gravitational backreaction should, in principle, be included when solving for the metric. We thus check the validity of the assumption of an \AdS spacetime in the presence of the \vev. The amount of backreaction is estimated from Einstein's equation, where the \AdSdpp cosmological constant term is
\begin{align}
    g_{MN} \Lambda_\textnormal{AdS} = -g_{MN}\frac{d(d-1)}{2 R^2} \ .
\end{align}
The assumption of an \AdS spacetime is valid as long as $\Lambda_\textnormal{AdS}$ is large compared to the scalar \vev contribution to  the classical stress tensor, $M^{1-d}_* T^{}_{MN}$, with 
\begin{align}
    T^{}_{MN}
    &=
    \left[
        -\eta_{MN} 
        \frac{d}{2}\left(\frac{d}{2}-\alpha\right)
        -\delta^z_{M} \delta^z_{N} \left(\frac{d}{2}-\alpha\right)^{\!2}
    \right]
    \frac{\vuv^2}{R^2} \left(\frac{z}{R}\right)^{d-2-2\alpha} \ 
    \label{eq:vev:stress:energy}
\end{align}
where $\eta_{MN}$ is the $(d+1)$-dimensional Minkowski metric. By na\"ive dimensional analysis we assume that $\vuv^2 \sim R^{1-d} \ll M_*^{d-1}$~\cite{Manohar:1983md,Georgi:1986kr}. The backreaction depends on the value of $\alpha$ relative to the dimension of the spacetime.
With these inputs, we show that one may always take the spacetime to be approximately \AdS near the \UV brane.

\paragraph{Range $\alpha\geq d/2$:}  
In this case, imposing the regularity condition $C=0$ on \eqref{eq:vev:boundary} means that $v(z)$ decreases rapidly enough that $T_{MN}$ is negligible for any $z>\zUV$.
The backreaction is thus neglible and the spacetime is \AdS  for any $z>\zUV$. Notice the regularity condition is equivalent to imposing a generic (not tuned) boundary condition on the \IR brane and taking  the limit where the brane position goes to $\zIR \to \infty$. 

\paragraph{Range $\alpha < d/2$:} 
In this case, the scalar \vev contribution to the stress-energy tensor $T^{}_{MN}$ grows with $z$. 
Consider the Minkowski component, $T^{}_{\mu\nu}$.
The scalar \vev and \AdS cosmological constant are of the same order of magnitude when $z$ satisfies
\begin{align}%
    \frac{d(d-1)}{2 R^2} 
    \sim  
    \frac{d}{2}
    \left( \frac{d}{2} - \alpha \right) 
    \frac{\vuv^2}{R^2} 
    \left( \frac{z}{R} \right)^{\!d-2\alpha} 
    M_*^{1-d} \ .
\end{align}
One can no longer ignore the gravitational backreaction due to the \vev for $z$ larger than this value.
In this case, we model the backreaction with an infrared brane located at $\zIR$ satisfying
\begin{align}
    \zIR^{d-2\alpha} 
    \sim 
    R^{d-2\alpha} 
    \frac{2(d-1)}{d-2\alpha}
    \frac{M^{d-1}_*}{\vuv^2} \ .
    \label{eq:zIR}
\end{align}
The region $\zUV < z < \zIR$ is approximately \AdS.
The \IR brane requires that the vev has a boundary condition at $\zIR$.
A generic boundary condition (either Dirichlet or Neumann) on the \IR brane naturally sets
\begin{align}
    B \sim C\left(\frac{\zIR}{R}\right)^{\!2\alpha} \ , \label{eq:CBrel}
\end{align}
up to $\mathcal O(1)$ prefactors. In this case, the $C$ term is small relative to the $B$ term in the $\zUV < z < \zIR$ region. Thus we can assume that the spacetime is \AdS in this region. We confirm the validity of this approximation in Appendix~\ref{se:softwall} using an explicit soft wall construction. In Section~\ref{sec:spectrum:and:self:energy} we show that the low-energy states of the spectrum are insensitive to this modeling.

\subsection{Boundary conditions and gauge symmetry}
\label{sec:BC:and:gauge}

Fields in a spacetime with boundary must have specified boundary conditions in order to have well-posed Sturm--Liouville wave equations. 
The boundary conditions for gauge fields must be compatible with the gauge symmetry of the system: they must be gauge invariant and uniquely project onto a physical field configuration~\cite{Vassilevich:1994cz,Vassilevich:1997iz}.

The $d$-vector $A_\mu$ and the scalar $A_z$ boundary conditions are compatible when one is Dirichlet and the other is Neumann~\cite{Csaki:2003dt,Cox:2019rro}.\footnote{We refer to any boundary condition that includes a first derivative as Neumann; this includes conditions that are sometimes referred to as Robin boundary conditions.} This restriction is necessary to avoid over-constraining the gauge parameter. For example, Ref.~\cite{Witten:2018lgb} shows that imposing the same boundary condition on $A_\mu$ and $A_z$ would force the Fadeev--Popov ghost field---the quantization of the gauge parameter---to simultaneously satisfy Dirichlet and Neumann boundary conditions. Because the ghost obeys a second-order wave equation, such a  condition is not consistent.

We choose a Neumann boundary condition for the $d$-vector $A_\mu$ and a Dirichlet boundary condition for the scalar $A_z$. This means that the $A_\mu$ component fluctuates on the brane while the $A_z$ component does not. 
The $A_\mu$ thus has an additional boundary degree of freedom, $A^\mu_\textnormal{UV} \equiv A^\mu(z=\zUV)$, that would be absent in the case of a Dirichlet boundary condition. This degree of freedom appears in an additional $\mathcal D A^\mu_\textnormal{UV}$ in the path integral measure.
The Neumann boundary condition is governed by  the variation of the action with respect to $A^\mu_\textnormal{UV}$.  The generic form of the boundary conditions for $A_\mu$ and $A_z$ is 
\begin{align}
    \left.
    \left(
       \eta^{\mu\nu}\partial_z - \mathcal B^{\mu\nu} 
    \right) A_\nu(x,z)
    \right|_{\zUV} &= 0 \,, 
    &
    \delta A_z(x,\zUV) &= 0    \,,
    \label{eq:A:BC:Neu:Dir}
\end{align}
where $\mathcal B^{\mu\nu}$ encodes brane-localized kinetic terms and is derived in \eqref{eq:Bmunu}. The $\partial_z$ factor comes from the surface term in the integration by parts of the bulk kinetic term.

The Neumann boundary condition for the $d$-vector $A_\mu$  in \eqref{eq:A:BC:Neu:Dir} 
implies that there is a separate $d$-dimensional gauge redundancy for $A^\mu_{\rm UV}$   with a $z$-independent gauge parameter $\Omega_\textnormal{UV}$. 
With our choice of boundary conditions, it follows that the gauge redundancies are
\begin{align}
    A_M(x,z)  &\to A_M(x,z)+\partial_M \gauge(x,z)\,,
    &
    A_\textnormal{UV}^\mu(x) &\to A_\textnormal{UV}^\mu(x) + \partial^\mu \gauge_\textnormal{UV}(x) 
    \,. 
    \label{eq:bulk:brane:gauge}
\end{align}
For the opposite choice of boundary conditions---Dirichlet $A_\mu$ and Neumann $A_z$---there is no residual gauge redundancy on the brane.

\subsection{Gauge Fixing}

Gauge redundancies must be fixed to project to physical field configurations when quantizing a theory. We use the Faddeev--Popov procedure to fix the redundancies in \eqref{eq:bulk:brane:gauge}. 
Additionally, both bulk and brane Lagrangians feature kinetic mixing between the $d$-vector $A_\mu$ and the scalars $A_z$ and $\pi$. We would like to remove these mixings so that we have separate kinetic operators for fields of different $d$-dimensional Lorentz representations.
This is conveniently realized by using versions of $R_\xi$ gauge-fixing functionals adapted to warped space. 
Similar approaches and related discussions can be found in Ref.~\cite{Cacciapaglia:2005pa,Barvinsky:2005qi,Gomes:2019xhu, Cheng:2023bcv, Muck:2001yv,Randall:2001gb}.

The bulk gauge kinetic term \eqref{eq:bulk:Lagrangian} contains a kinetic mixing between the $d$-vector, $A_\mu$, and the scalar $A_z$.  The bulk scalar kinetic term also introduces a kinetic mixing between $A_\mu$ and the derivative of the Goldstone $\partial_\mu \pi$ in the presence of a \vev. 
The bulk gauge fixing action that cancels these kinetic mixings is
    \begin{align}
        S_{\textnormal{fix}}^{\textnormal{Bulk}}
        =
        &
        \frac{-1}{{\g}^2} 
        \int 
        \Ddppx\,
        \left(\frac{R}{z}\right)^{\!d-3}\frac{1}{2\xi}
        \left[
            \partial_\mu A^\mu
            -
            \xi\left( 
                z^{d-3} \partial_z \left(\frac{A_z}{z^{d-3}}\right)
                -
                \left(\frac{R}{z}\right)^{\!2} {\g}^2 v(z)\, \pi
            \right)
        \right]^2 \ ,
        \label{eq:SGFBulk}
    \end{align}
where $\xi$ is the bulk $R_\xi$ gauge  parameter. 

In the brane-localized piece of the action, inserting the Higgs \vev in the brane-localized kinetic term for the scalar \eqref{eq:LUV}  generates a mixing term between $A^\mu$ and the Goldstone $\partial_\mu\pi$. 
Furthermore, integrating the bulk action by parts induces a surface term that is a brane-localized mixing between $A^\mu$ and $\partial_\mu A_z$. 
The brane gauge fixing action that cancels these kinetic mixings is
\begin{align}
        S_{\textnormal{fix}}^{\textnormal{UV}}
        =&
        -
        \int_\textnormal{UV} 
        \Ddx\,
        \frac{1}{2\xi_{\textnormal{UV}}R^{d-4}}
        \left[
            \partial_\mu A^\mu
            +
            \xi_{\textnormal{UV}}
            R^{d-4}
            \left(
                c_\textnormal{UV}\, \vuv R\, \pi
                -
                \frac{1}{g^2}
                \frac{R}{z}
                A_z
            \right)        
        \right]^2
        \, ,
        \label{eq:SGFUV}
    \end{align}
where $\xi_\textnormal{UV}$ is the \UV brane $R_\xi$ gauge parameter.

\subsection{Effective Theory Cutoff}
\label{sec:cutfoffscales}

The interacting $(d+1)$-dimensional theory is understood to be a long-distance effective field theory that is valid for distances longer than a $(d+1)$-dimensional cutoff scale, $\Delta x \sim {\Lambda^{-1}_{(d+1)}} $. The ultimate cutoff of the $(d+1)$-dimensional theory is proportional to the $(d+1)$-dimensional Planck mass. However, in this paper we assume a much lower effective theory cutoff at the \AdS curvature:
\begin{align}
\Lambda_{d+1} \sim\frac{1}{R}\,. \label{eq:cutoff}     
\end{align}
This is the cutoff for the validity of the \AdSCFT correspondence~\cite{Aharony:1999ti}. 
This corresponds to a four-momentum cutoff $p_\textnormal{cutoff}\sim \zUV^{-1}$ in the boundary correlation functions (see e.g. \cite{Fichet:2021xfn}).

\section{Green's Functions and Boundary Effective Action}
\label{sec:propagators}

We present the general structure of the Green's function equations and boundary effective actions in our model.  Starting from the full \UU action \eqref{eq:full_action}, we separate the $d$-dimensional vector sector formed by the gauge field components parallel to the brane, $A_\mu$, and the scalar sector composed of the bulk Higgs and $A_z$. 
We present the Green's function equations for the vector sector in Section~\ref{sec:vector_prop} and discuss the associated scalar sectors in Section~\ref{sec:scalar_prop}. The holographic basis and the  boundary effective action are introduced in Sections~\ref{sec:hol_basis} and \ref{sec:hol_action}.

\subsection{Vector Green's Function Equations}
\label{sec:vector_prop}

\paragraph{Green's function equation} We focus on the  $A_\mu$ sector from the \UU action \eqref{eq:full_action}. 
Upon integration by parts in the Minkowski directions, the quadratic terms in \eqref{eq:full_action}  are 
\begin{align}
    S
    \supset
    \frac{1}{2{\g}^2}
    \int 
    \Ddppx
    \, 
    A_\mu \mathcal{O}^{\mu \nu}A_\nu
    + 
    \frac{1}{2{\g}^2}
    \int_\textnormal{UV} 
    \Ddx
    \, 
    A_\mu\left(
        \mathcal{B}^{\mu \nu}-\eta^{\mu \nu}\partial_z
        \right)A_\nu
        \label{eq:S_quad}
\end{align}
where all lowercase Greek indices are contracted according to the $d$-dimensional Minkowski metric. The bulk and brane kinetic operators are, respectively,
\begin{subequations}
\label{eq:kinetic:operators}
\begin{align}
    \mathcal{O}^{\mu \nu}
    &=
    \left(\frac{R}{z}\right)^{\!d-3}
    \left[
        \eta^{\mu \nu} \partial^2
        -   \left( 1 - \frac{1}{\xi} \right)
            \partial^\mu \partial^\nu
        - \eta^{\mu \nu} z^{d-3} \partial_z 
            \left( \frac{ 1 }{ z^{d-3} } \partial_z \right)
        + \left(\frac{{\g}v(z) R}{z}\right)^{\!2}
    \right] 
    \label{eq:O}
    \\
    \mathcal{B}^{\mu \nu}
    &
    =
    \auv R
    \left[
    \eta^{\mu \nu} \partial^2
    - 
    \left( 
        1 - \frac{ {\g}^2 }{ \xi_{\textnormal{UV}} \auv R^{d-3} } 
    \right) 
    \partial^\mu \partial^\nu 
    +
    \frac{\cuv}{\auv} \mA ^2
    \right]
    \label{eq:Bmunu}
    \ , 
\end{align}
\end{subequations}
where we define the gauge field {effective} bulk mass on the \UV brane, 
$
    \mA ^2
    =
    {\g}^2 \vuv^2
$.
The gauge field homogeneous equation of motion and Neumann boundary condition are
    \begin{align}
        \mathcal{O}^{\mu \nu} A_\nu & =0
        &
        \left. \left(
            \eta^{\mu \nu}\partial_z-\mathcal{B}^{\mu\nu}
            \right)
        A_\nu \right|_{\zUV}
       =
       0
       \, .  
       \label{eq:BoundaryA}
    \end{align}
The  $A_\mu$ propagator from $x$ to $x'$ is the Green's function for the bulk operator that satisfies
\begin{align}
    \mathcal{O}^{\mu \rho}
    \langle A_\rho(x)\, A_\nu (x')\rangle
    =
    i {\g}^2 \,
    \delta^\mu_\nu \;
    \delta^{(d+1)}(x-x') \ .
\end{align}
Introduce the Fourier transform along the $d$ Minkowski spactime directions,
\begin{align}
    A_M(x,z) 
    =
    \int 
    \Ddp
    \,
    \e^{i p_\mu x^\mu}
    A_M(p,z) 
    \ ,
\end{align}
where we define $p^2=\eta_{\mu\nu}p^\mu p^\nu $. 
With our metric signature, $p^2<0$ for spacelike momentum and $p^2>0$ for timelike momentum.
In a free theory, $p^2$ is made slightly complex to resolve the  non-analyticity in Green's functions with timelike momentum. This corresponds to the inclusion of an infinitesimal imaginary shift $p^2 \to p^2+i\varepsilon$ with $\varepsilon\rightarrow 0^+$. This prescription imposes causality and  defines the Feynman propagator. We leave this $i\varepsilon$ shift implicit in the remainder of this paper. 

\paragraph{Decomposition by polarization} The vector field decomposes into transverse and longitudinal {components} 
\begin{align}
    A_\mu (p, z)
    =
    P^\textnormal{T}_{\mu \nu}
    A^\nu_\textnormal{T}(p,z)
    +
    P^\textnormal{L}_{\mu \nu}
    A^\nu_\textnormal{L} (p, z)
    \, ,
    \label{eq:AL:AT}
\end{align}
where we define the projection operators onto spaces perpendicular and parallel to momentum,
\begin{align}
    P^\textnormal{T}_{\mu \nu} 
    &=   \eta_{\mu \nu} - 
        \frac{ p_\mu p_\nu }{ p^2 } 
    &
    P^\textnormal{L}_{\mu \nu} 
    & = 
    \frac{ p_\mu p_\nu }{p^2 }
    \, .    
\end{align}
We use these projection operators to decompose the kinetic operators \eqref{eq:kinetic:operators}:
\begin{align}
    \mathcal{O}_{\mu\nu} 
    &= \mathcal{O}_\textnormal{T} P^\textnormal{T}_{\mu \nu}  
        + 
    \mathcal{O}_\textnormal{L} P^\textnormal{L}_{\mu \nu}
    &
    \mathcal{B}_{\mu\nu} &= 
    \mathcal{B}_\textnormal{T} P^\textnormal{T}_{\mu \nu}  
    + 
    \mathcal{B}_\textnormal{L} P^\textnormal{L}_{\mu \nu}
    \label{eq:tensor:to:scalar:operators}
    \,.
\end{align}
This defines the transverse and longitudinal components of the bulk equation of motion and of the boundary condition. 
The equations in \eqref{eq:BoundaryA} become 
\begin{align}
    \mathcal O_\textnormal{X} A^\mu_X&=0
    &
    \left.
    \left(
    \partial_z-
        \mathcal B_X
    \right) A^\mu_X
    \right|_{\zUV}
    =0 \ ,
    \label{eq:BC:on:gauge:no:indices}
\end{align}
with $X\in \{\textnormal{T, L}\}$. 
The expressions for the transverse kinetic operators are 
\begin{subequations}
\label{eq:tranverse:kinetic:operators}
\begin{align}
    \mathcal{O}_\textnormal{T}(p^2) 
    &= 
    \left(\frac{R}{z}\right)^{d-3} 
    \left[
        - \partial_z^2 
        + \frac{d-3}{z}\partial_z
         -p^2
        +\mA ^2 
            \left(\frac{z}{R}\right)^{d-2-2\alpha}
    \right]
    \label{eq:EOM:AT}
    \\
 \frac{ \mathcal{B}_\textnormal{T} (p^2)}{R}
    &=
       - \auv p^2
        +
        \cuv \mA ^2
    \, . 
    \label{eq:B_op}
\end{align}
\end{subequations}
The longitudinal operators are simply related to the transverse ones by a replacement of the $d$-momentum:
\begin{align}
    \mathcal{O}_\textnormal{L}(p^2) 
    &=  
    \mathcal{O}_\textnormal{T}\!
    \left( \frac{p^2}{\xi} \right)
    &
    \mathcal{B}_\textnormal{L} (p^2)
    &= 
    \mathcal{B}_\textnormal{T}\! 
    \left( 
        \frac{ g^2\,p^2 }{ \auv \xi_\textnormal{UV} R^{d-3} }
    \right)
    \, . 
    \label{eq:EOM:and:B_op:L}
\end{align}
This result is consistent with \cite{Randall:2001gb}. 

\paragraph{Propagator}
The propagator decomposes into transverse and longitudinal pieces,
\begin{align}
   \langle A_\mu(p,z) \, A_\nu(-p,z')\rangle
    &=
     i
    P^\textnormal{T}_{\mu \nu}
    G^\textnormal{T}_p(z,z')
    +i
    P^\textnormal{L}_{\mu \nu}
    G^\textnormal{L}_p(z,z')
    \ ,
    \label{eq:propdcomp}
\end{align}
where the scalar propagators $G_p^X(z,z')$ satisfy 
\begin{align}
    \mathcal{O}_{\!X} G^X_P &= g^2\delta(z-z')
    \label{eq:EOM:G}
\end{align}
and the Neumann boundary condition \eqref{eq:BoundaryA}. 
We can see that decomposing the vector field into transverse and longitudinal polarizations reduces the Green's function equation and boundary conditions to those of  scalar propagators $G_p^X$. In turn, $G_p^\textnormal{T,L}$ are related to each other by simple replacements of parameters.

\subsection{Scalar Green's Function Equations}
\label{sec:scalar_prop}

The spin-zero sector contains the gauge  pseudoscalar $A_z$, the pseudoscalar Goldstone $\pi$, and the scalar radial mode $h$. In what follows, it is convenient to define a dimensionless Goldstone field
\begin{align}
    a(x,z)  \equiv \frac{\pi(x,z)}{v(z)} \ .
    \label{eq:a:def:pi:v}
\end{align}
Under a gauge transformation this field transforms as a shift, $a\to a+\gauge$.

\paragraph{Scalar.}
The equation of motion for $h$ follows directly from that of the Higgs, \eqref{eq:Phi:EOM}:
\begin{align}
    \left(\frac{R}{z}\right)^{\!d-1}
    \left[
        p^2 
        +
        z^{d-1} \partial_z
        \left( 
            \frac{1}{ z^{d-1} \partial_z }
        \right)
    -
    \left( \frac{ \mu R }{ z } \right)^{\!2}
    \right]
    h
    =0
    \label{eq:h:EOM}
    \, .
\end{align}
The boundary condition for the radial mode is
\begin{align}
    \left.
    \left(
        \partial_z 
        +
        c_{\textnormal{UV}} p^2 R
        - m_h^2 R 
    \right) h
    -
    3 \lambda_{\textnormal{UV}}v_\textnormal{UV} R^{d-2}\, h^2
    -
    \lambda_{\textnormal{UV}}R^{d-2}\, h^3
    \right|_{\zUV}
    =
    0 \ 
\end{align}
where we identify the brane-localized mass $m_h^2 = 2\vuv^2 \lambda_\textnormal{UV}R^{d-3}$. 
We disregard the radial mode for the remainder of this work since it carries no further implication on the gauge sector.

\paragraph{Pseudoscalars.}  
The $\pi$ and $A_z$ fields mix with each other in the bulk quadratic action.
Varying the action with respect to $\pi$ and $A_z$ give the following homogeneous equations of motion:
\begin{subequations}
\label{eq:pseudscalar:EOM}
\begin{align}
    0&=
    v(z)^2\, p^2 \,a(p,z)
    +
    \left( \frac{z}{R} \right)^{\!d-1} 
   \partial_z 
    \left[
        \left( \frac{R}{z} \right)^{\!d-1}
        \!
        v(z)^2 \,
        \chi(p,z)
    \right]
    +
    \xi v(z)^2 \,
    \Theta(p,z)
    \label{eq:piEOM}
    \\ 
    0&=
    p^2 A_z(p,z)
    + {\g}^2
    \left( \frac{R}{z} \right)^{\!2}
    v(z)^2\,
    \chi(p,z)
    +
    \xi 
   \partial_z
    \Theta(p,z)
    \ ,
   \label{eq:A5EOM}
   \, 
\end{align}
\end{subequations}
where we define the following combinations of the Goldstone $a$ and the gauge scalar $A_z$:
\begin{align}%
    \chi(p,z) 
    &= \partial_z a-A_z 
    &
    \Theta(p,z)
    &= z^{d-3} \partial_z 
        \left( 
            \frac{A_z}{z^{d-3}}  \right)
            - \g^2 \left( \frac{R}{z} \right)^{\!2} v(z)^2\, a  
        \, .
\label{eq:chiThetadef}
\end{align}
The gauge-invariant pseudoscalar $\chi$ is decoupled from the vector sector.  Like the radial mode, we disregard $\chi$ for the remainder of this study.

This combination of fields allows us to decouple the equations of motion \eqref{eq:pseudscalar:EOM},\footnote{%
    When $\alpha={d}/{2}$, the \UU breaking \vev is constant, $v=\vuv$, and both equations of motion may be solved exactly. The two solutions for $\chi$ are 
    $z^{d/2} \textnormal{B}_\nu(\sqrt{-p^2}z) $ where $\textnormal{B}_\nu$ are the modified Bessel functions of the second kind, $\textnormal{B}_\nu \in \left\{I_\nu, J_\nu\right\}$. The index $\nu$ depends on the \vev: $\nu^2 = (d/2-1)^2 + g^2\vuv^2 R^2$. The $\Theta$ solutions are related to those for $\chi$ by $\sqrt{-p^2}\to \sqrt{-p^2/\xi}$ in the arguments.  \label{foot:d2:exact}  
}
\begin{align}%
    \partial_z 
    \left[
        \frac{z^{d-1}}{v(z)^2}
        \partial_z
            \left(
                \frac{v(z)^2}{z^{d-1}} \chi 
            \right) 
        \right]
    +   p^2 \chi 
    -   g^2 v(z)^2 
        \left( \frac{R}{z} \right)^{\!2}
         \,\chi 
    &=0  
    \label{eq:EOMchi}
    \\
    z^{d-3} \partial_z
    \left(
        \frac{1}{z^{d-3}} \partial_z \Theta
    \right) 
    + \frac{1}{\xi} p^2 \Theta
    -  g^2 v(z)^2 
        \left( \frac{R}{z} \right)^{\!2}
        \Theta
    &=0 \label{eq:EOMTheta}
    \ .
\end{align}
These results are consistent with those of \cite{Cox:2019rro} upon appropriate matching of conventions. We recognize that $\Theta$ is the same linear combination of pseudoscalars that mix with $\partial_\mu A^\mu$ upon integrating the bulk kinetic term by parts. We then chose a bulk $R_\xi$ gauge fixing \eqref{eq:SGFBulk} to cancel this mixing. We show in Section~\ref{sec:GET} that $\Theta$ is identified with the longitudinal mode of $A_\mu$ in the unitarity limit.

We now turn to the boundary conditions of the pseudoscalar sector. 
In Section~\ref{sec:BC:and:gauge} we specified Dirichet and Neumann boundary conditions for $A_z$ and $a$ respectively.  Varying the action with respect to $\delta a$ gives
\begin{align}
    \left.
    \left[
        \left( \partial_z a - A_z \right)
        + c_{\textnormal{UV}} R\, p^2 a
        + c_{\textnormal{UV}} \xi_{\textnormal{UV}}
            \left( \frac{R^{d-3} }{ {\g}^2} \right)
            \left( A_z-{\g}^2 c_\textnormal{UV} R v^2 a \right)
    \right] \delta a
    \right|_{z_\textnormal{UV}}
        =
        0 \, . 
        \label{eq:BC_var_a}    
\end{align}
The Neumann boundary condition for $a$ implies that the bracket vanishes.  
The boundary condition $\delta A_z=0$, permits $A_z$ to be non-zero on the brane and only imposes that variations of $A_z$ vanish on the brane. 
{This means that some work is needed to specify a second pseudoscalar boundary condition. One hint is that} \eqref{eq:BC_var_a}  implies a boundary condition on the gauge transformation parameter,
\begin{align}
    \partial_z \gauge - g^2R 
    \left.\left( 
        c_\textnormal{UV} \vuv^2 
        - \frac{ p^2 }{ \xi_\textnormal{UV} R^{d-3} } 
    \right)\gauge\right|_{\zUV}
    &=0 \
    \,. \label{eq:gauge_BC}    
\end{align}
Since the gauge parameter cannot not simultaneously satisfy two different boundary conditions, the only consistent possibility is to split \eqref{eq:BC_var_a} into two equations:
\begin{align}
    \left.\partial_z a - 
    A_z \right|_{\zUV}
    &=0
    &
    \left.
    A_z - g^2 R 
    \left(
        c_{\textnormal{UV}} v^2 
        - 
        \frac{ p^2 }{ \xi_{\textnormal{UV}} R^{d-3} }
    \right) a
    \right|_{z_\textnormal{UV}}
    &=0 \,.
    \label{eq:BC_pseudoscalar}
\end{align}
The first boundary condition is gauge invariant and thus does not constrain $\Omega$, while the second boundary condition yields the same constraint on $\Omega$ as \eqref{eq:gauge_BC}. One further consistency check of this split is that \eqref{eq:gauge_BC} is the same boundary condition on the gauge transformation parameter as one derives from the longitudinal mode of $A_\mu$.

\subsection{The Holographic Basis}
\label{sec:hol_basis}

In a quantum field theory on a spacetime with boundary, it is useful to separate a quantum field's bulk and boundary degrees of freedom using a \textit{holographic basis}, see {e.g.} Refs.~\cite{Batell:2007jv, Fichet:2021xfn}. 
In this basis, the gauge field is
\begin{align}
    A^M(p;z) &= A^M_\textnormal{UV}(p)\,K(p;z) + A^M_\textnormal{D}(p;z) 
    \ .    
    \label{eq:hol_basis} 
\end{align}
The first term on the right-hand side contains the boundary degree of freedom $A_\textnormal{UV}(p)$. The profile $K(p;z)$ can be chosen to be the brane-to-bulk propagator. The second term on the right-hand side  is a bulk component that vanishes on the boundary, $A_\textnormal{D}(p;\zUV)=0$.%
\footnote{
    The holographic decomposition holds for bulk fields with any boundary condition. A Neumann (Dirichlet) boundary condition means that $A_\text{UV}$ does (not) fluctuate on the \UV brane.
    Though the subscript \acro{D} for `Dirichlet' of the bulk component is a common notation in the holographic decomposition, we stress that the  condition on $A_\textnormal{D}$ is distinct from the Dirichlet boundary condition, e.g.~\eqref{eq:A:BC:Neu:Dir}, which is a condition on the \emph{fluctuations} of the field. 
    }  
We assume the analogous decomposition for the pseudoscalar fields, $a$ and $\Theta$.
Introduce $G_\textnormal{N}(p;z,z')$ as the bulk Feynman propagator  satisfying Neumann boundary conditions~\cite{Ponton:2012bi}.  The brane-to-bulk propagator is defined as the amputated bulk-to-bulk Neumann propagator taken on the boundary 
\begin{align}
    K(p;z) &=
    \frac{ G_\textnormal{N}(p;z_\textnormal{UV},z) 
        }{ G_\textnormal{N}(p;z_\textnormal{UV},z_\textnormal{UV}) 
        } \ .
    \label{eq:K:p:z}
\end{align}
 In this basis the quadratic action is diagonal~\cite{Fichet:2021xfn}.

\subsection{Boundary Effective Action } 
\label{sec:hol_action}

Integrating out the bulk degrees of freedom---$A^M_{\rm D}$, $a_{\rm D}$, etc.---in the partition function yields a boundary quantum effective action $\Gamma$ that depends on the classical values of the boundary degrees of freedom,  $\Gamma\left[A_\textnormal{UV}, a_\textnormal{UV},\ldots\right]$. For the main part of this work we focus on the tree-level, quadratic piece of this effective action, $ \left.\Gamma^{(2)}\right|_\textnormal{tree}$. The relations shown in this section are  general features of the holographic formalism, similar ones hold for any manifold with boundary.\,\footnote{See {e.g.}~\cite{Ponton:2012bi} for a warped background, \cite{Fichet:2021xfn} for a generic manifold with boundary.}

\subsubsection*{Gauge Boson Boundary Effective Action}

At tree-level, the boundary effective action is evaluated by plugging in the bulk equation of motion into the fundamental action, $S$.
The quadratic piece of the gauge boson tree-level, boundary effective action  is
\begin{align}
    \left.  \Gamma_A^{(2)}\right|_\textnormal{tree}
        &=
       \frac{1}{2{\g}^2} 
       \int 
       \Ddp
       \sum_{X=\textnormal{T},\textnormal{L}}
       A^\mu_\textnormal{UV} (p)
       \left[
            \mathcal{B}_X (p^2)+\Sigma_X (p^2)
        \right]
       P^X_{\mu \nu}
       A^\nu_\textnormal{UV} (-p)
       \ .
        \label{eq:SboundAmu}
    \end{align}
The index $X\in\{\textnormal{T},\textnormal{L}\}$ denotes the transverse and longitudinal components of $A^\mu$. $\mathcal{B}_X$ is the quadratic boundary-localized action in \eqref{eq:tensor:to:scalar:operators}. $\Sigma_\textnormal{X}$ is the \emph{holographic self-energy}, the normal derivative of the boundary-to-bulk propagator, see {e.g.}~\cite{Nastase:2007kj, Ponton:2012bi}, 
\begin{align}
   \Sigma_{X} (p^2) = - \left.\partial_z \log K_{\!X}(p;z)\right|_{\zUV} \ .
   \label{eq:holoselfenergy}
\end{align}
$\Sigma_\textnormal{X}$ is the inverse of the brane-to-brane propagator in the limit of no boundary operator, $\mathcal{B}\to 0$. The longitudinal and transverse self-energies are related by
\begin{align}
    \Sigma_\textnormal{L}(p^2) &= 
    \Sigma_\textnormal{T}\left( \frac{p^2}{\xi} \right) 
    \,. 
    \label{eq:SigmaTL_rel}
\end{align}
This follows from the analogous relations for the bulk and boundary operators, \eqref{eq:EOM:and:B_op:L}. The boundary operators are analytic in $p$, while the $\Sigma(p^2)$ pieces are, in general, non-analytic in $p$. The brane-to-brane propagator is 
\begin{align}
    G_{X}(p;\zUV, \zUV)
    &=
    \frac{{\g}^2}{\mathcal{B}_{X}(p^2)+\Sigma_{X}(p^2)}\,.
    \label{eq:b2b_G}  
\end{align}
$G_\textnormal{X}^{-1}$ can be obtained from the boundary effective action $\Gamma_A$ by taking two functional derivatives with respect to $A_\textnormal{UV}$.

\subsubsection*{Boundary-Localized Limit}

In the brane-localized limit $\auv,\,\cuv \gg 1$ the effective boundary action \eqref{eq:SboundAmu} is
\begin{align}
    \left. \Gamma_A^{(2)} \right|_\textnormal{tree}&\;
    \xrightarrow[\auv,\,\cuv \gg 1]{}
    \;
    \frac{-R}{ 2 {\g}^2 } 
    \int 
    \Ddp
    \sum_{ X=\textnormal{T}, \textnormal{L} }
    A^\mu_\textnormal{UV} (p)
    \left(
        \auv p^2
        -
        \cuv \mA ^2
    \right) P^X_{\mu \nu}
   A^\nu_\textnormal{UV} (-p) \ .
\end{align}
This action describes a  gauge field with squared mass  $\left(\cuv/\auv\right)\mA ^2$ in $d$-dimensional flat space. 

\subsubsection*{Goldstone Boundary Action}
\label{sec:hol_action_GB}

Denote the tree-level quadratic piece of the boundary effective action for the  Goldstone degree of freedom by $\left. \Gamma^{(2)}_a \right|_\textnormal{tree}$.  {In anticipation of the Goldstone equivalence theorem,} we consider this sector in the $p\gg {\g}\vuv $ limit. This limit is conveniently obtained by work{ing} perturbatively in ${\g}$.  With our gauge field normalization, the bulk value of $A_z$ is of order $\g^2$~\cite{Cox:2019rro}. 
To leading order in ${\g}$, the Goldstone boson equation of motion is,
\begin{align}
    -\left(\frac{R}{z}\right)^{d-1} v^2 \partial^2 a
    +
    \partial_z \left[\left(\frac{R}{z}\right)^{d-1}v^2\partial_z a\right]
    =  0
    \, .
    \label{eq:piEOMLO}
\end{align}
The solutions and their $|p|z \ll 1$ asymptotics are $z^{\alpha}I_{\alpha}(\sqrt{-p^2} z ) \sim z^{2\alpha}$ and $z^{\alpha}K_{\alpha}(\sqrt{-p^2} z ) \sim z^0$. The asymptotic behaviors of $\pi(x,z)$ is thus $z^{{d}/{2}\pm\alpha}$.   
From \eqref{eq:holoselfenergy}, the holographic self-energy of $a$ is
\begin{align}
    \Sigma_a(p) 
    & =  
    \sqrt{-p^2} 
    \frac{K_{\alpha-1}(\sqrt{-p^2}\zUV)}{K_{\alpha}(\sqrt{-p^2}\zUV)} 
    \approx 
    \frac{p^2 \zUV }{(2-2\alpha)}  
    +\frac{2}{\zUV}
        \frac{\Gamma(1-\alpha)}{\Gamma(\alpha)}
        \left(\frac{-p^2\zUV^2}{4}\right)^{\alpha} \ .
 \label{eq:Sigmaa}
\end{align}
The Goldstone holographic self energy has no mass term in agreement with the shift symmetry of the Goldstone mode.

The Goldstone boson effective boundary action is
\begin{align}
    \Gamma_a
    &=
    - \frac{ v_\textnormal{UV}^2 }{2} 
    \int 
    \Ddp
    a_\textnormal{UV}(p)
    \left[ 
        \mathcal{B}_a(p^2) + \Sigma_a(p^2)
   \right]  
   a_\textnormal{UV}(-p)
   \label{eq:SboundGB}
\end{align}
with brane-to-brane propagator and corresponding boundary term
\begin{align}%
G_a(p;z_\textnormal{UV},z_\textnormal{UV})
    &=
 \frac{v_\textnormal{UV}^{-2}}{\mathcal{B}_a(p^2)+\Sigma_a(p^2)}
 &
\mathcal{B}_a(p^2) &= \frac{- g^2 p^2}{\xi_\textnormal{UV} R^{d-2}} + c_\textnormal{UV} R \,g^2\vuv^2 \,. 
\label{eq:B_a:b2b_a}
\end{align}
Notice that $\mathcal{B}_a$ is proportional to $\mathcal{B}_\textnormal{L}$, see~\eqref{eq:EOM:and:B_op:L}.

\section{The Goldstone Equivalence Theorem in AdS}
\label{sec:GET}

Spontaneously broken gauge theories in flat space feature a Goldstone boson equivalence theorem. This equates the $S$-matrix element for the emission of a longitudinally-polarized gauge field with that of the emission of a Goldstone boson $\pi$ in the limit of large momentum:
\begin{align}
    \varepsilon^\textnormal{L}_\mu(\p) 
    \mathcal{M}_A^\mu(\p) 
    \xrightarrow[p^2\gg m^2]{}%
    \mathcal{M}_\pi(\p)
    \ .
 \label{eq:GET_flat_simple}    
\end{align}
In this section we show the analogous features in our \AdS model.

\subsection{Review of Minkowski Derivation}
\label{sec:cutting:argument:for:GET}

A convenient way to derive the equivalence theorem in Minkowski space is to perform unitary cuts on the internal gauge and Goldstone lines of the appropriate diagrams~\cite[Eq.~(21.68)]{Peskin:1995ev}. In \tHF gauge, the numerator of the gauge boson propagator is proportional to
\begin{align}
    -g_{\mu\nu} 
    = 
    \sum_{\textnormal{polarizations, }s} 
    \varepsilon_\mu^s(\p) \, \varepsilon_\nu^s(\p) 
    - 
    \frac{\p_\mu \p_\nu}{\mA ^2} \ ,
\end{align}
where the second term is an unphysical timelike polarization that is canceled by a diagram where the gauge boson internal line is replaced with that of a Goldstone boson, $\pi$. Performing a unitarity cut along these internal lines implies 
\begin{align}
    -\left|\frac{\k_\mu}{m}\mathcal{M}_A^\mu(\p)\right|^2
    +\left|\mathcal{M}_\pi(\p)\right|^2
    =0 \ .
\end{align}
At large momentum, the longitudinal polarization vector approaches%
\begin{align}
    \varepsilon^\textnormal{L}_\mu(\p) 
    \xrightarrow[\p^2\gg m^2]{}%
    \frac{\p_\mu}{m}
    \ .
    \label{eq:GET:long:polz}
\end{align}
With this identification, the unitarity cut equation implies the equivalence theorem \eqref{eq:GET_flat_simple}.

\subsection{Two Goldstone Equivalence Theorems}
\label{sec:GET_AdS}

{In order to demonstrate the Goldstone equivalence theorem in \AdS, we must identify which degrees of freedom play the role of the Goldstone bosons and to which types of amplitudes the theorem should apply. }
In extra dimensional theories, the longitudinal polarization of the massive modes of a gauge boson can be understood as a consequence of `eating' the corresponding $A_z$ modes~\cite{Chivukula:2001esy}. On the other hand, the bulk scalar $\Phi$ furnishes a bulk Goldstone boson $\pi$ that also contributes to the longitudinal polarization of $A_\mu$. The Goldstone equivalence theorem thus involves a linear combination of $A_z$ and $\pi$.

{We identify the appropriate linear combination through the mixing terms between $A_\mu$ and pseudoscalar fields. We chose our $R_\xi$ gauge fixing to remove these terms in \eqref{eq:SGFBulk} and \eqref{eq:SGFUV}. Because there are both bulk and brane mixing terms, we expect separate bulk and boundary equivalence theorems that apply to different types of amplitudes.}

Our strategy to derive these Goldstone equivalence theorems is to mimic a flat space argument. This argument invokes the unitarity of the $S$-matrix through Cutkosky cuts of the intermediate $A_\mu$ propagator. To disentangle the bulk and brane effects, we decompose the Neumann propagator using the Neumann--Dirichlet identity
\begin{align}
    G_\textnormal{N}(\p;z,z') 
    &=  G_\textnormal{D}(\p;z,z') + K(\p;z) G_\textnormal{UV}(\p) K(\p;z') 
    \ ,
\end{align}
where $G_{\textnormal{N}, \textnormal{D}}$ are the Neumann/Dirichlet  propagators and $G_\textnormal{UV}(\p) = G_\textnormal{N}( \p; z_\textnormal{UV}, z_\textnormal{UV})$ 
is the brane-to-brane propagator and $K(\p;z)$ is the amputated boundary-to bulk propagator defined in \eqref{eq:K:p:z}. 
{We represent the Feynamn diagrams in our \AdS background using a Witten diagram notation where the circle represents the \UV brane, i.e.~the regulated \AdS boundary.}
A bulk diagram then decomposes into two contributions:
\begin{align}
    \vcenter{
        \hbox{\includegraphics[width=.15\textwidth]{{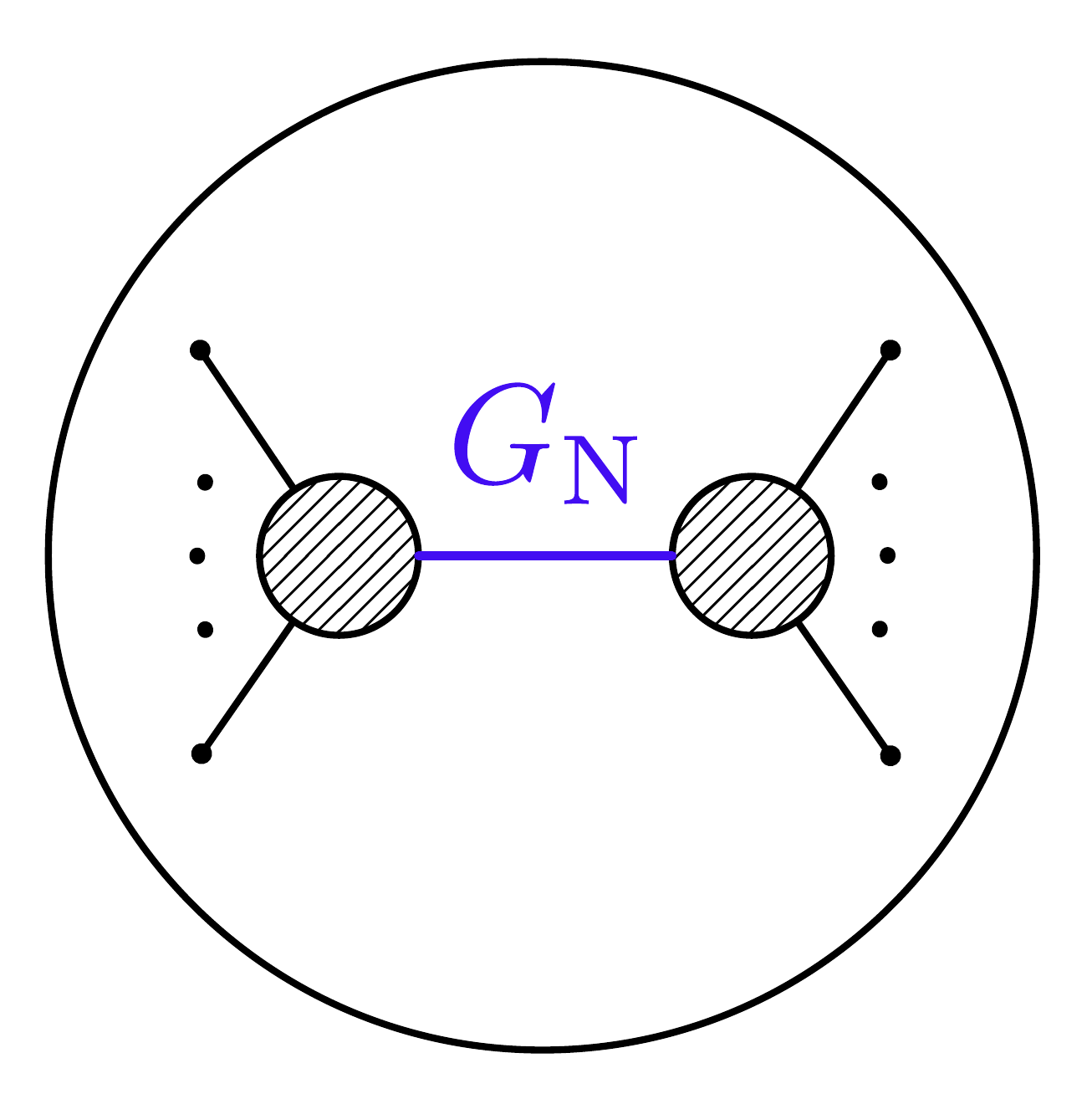}}}
        }
    &=
    \vcenter{
        \hbox{\includegraphics[width=.15\textwidth]{{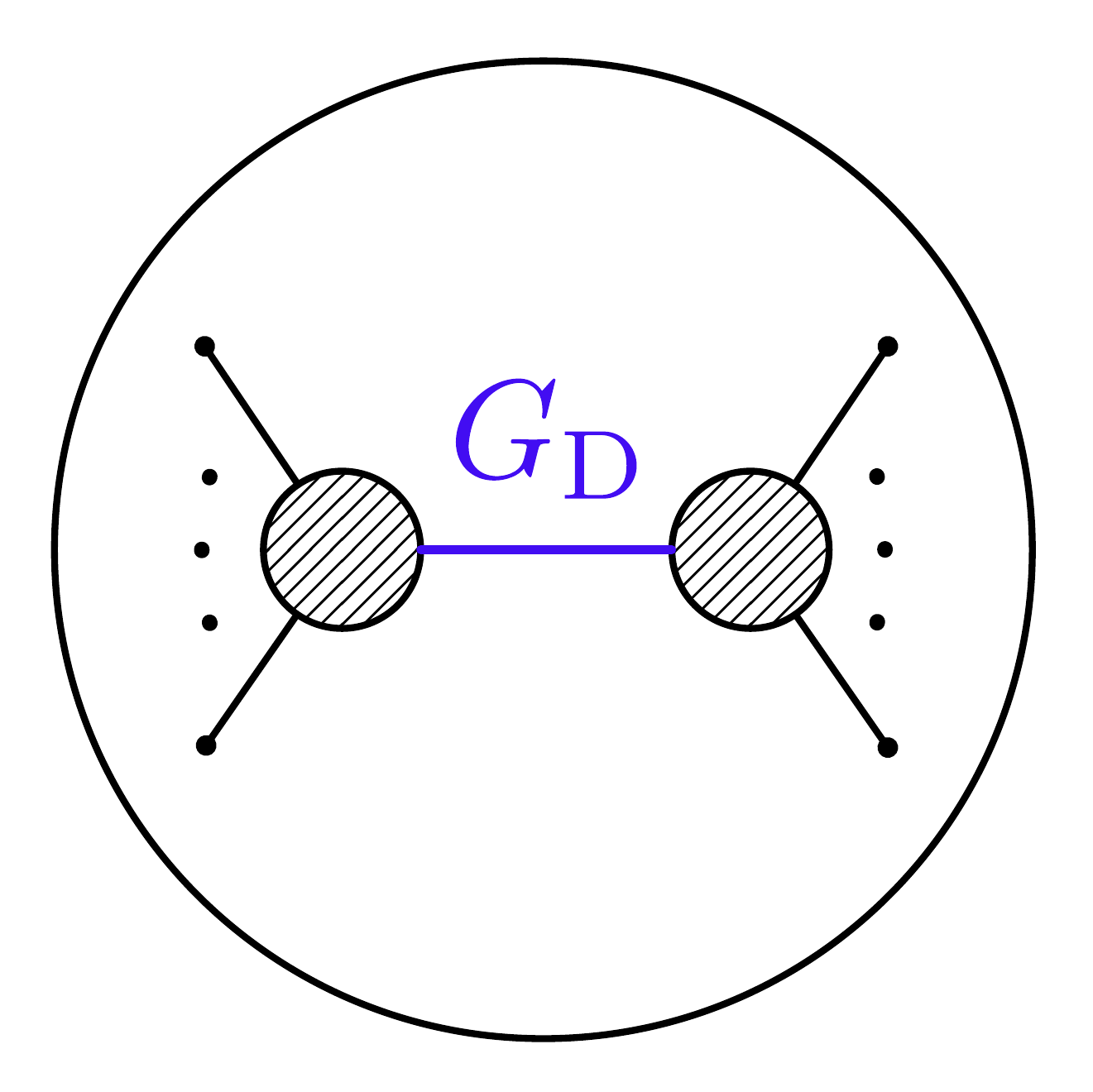}}}
        }
    +
    \vcenter{
        \hbox{\includegraphics[width=.15\textwidth]{{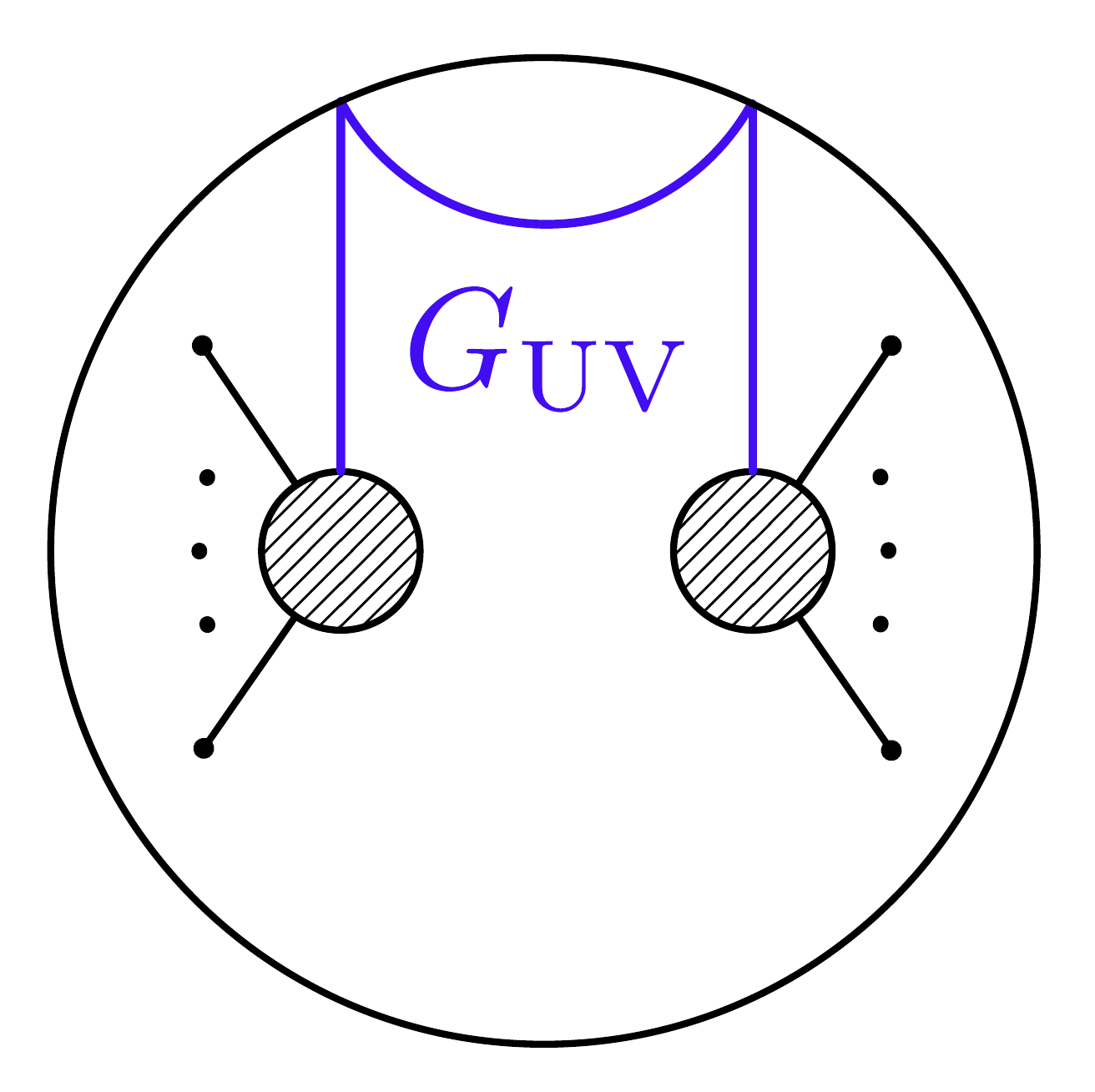}}}
        }
    \ . 
    \label{eq:GEB:sum}
\end{align}
The Minkowski space derivation invites us to make a unitarity cut on the internal line.  The cut applies to each diagrams on the right-hand side.  We analyze the two cut diagrams separately in Sections~\ref{sec:GET_bulk} and \ref{sec:GET_boundary}. They  respectively lead us to a bulk and a boundary version of the equivalence theorem.

\subsection{Bulk Equivalence Theorem}
\label{sec:GET_bulk}

We cut the Dirichlet internal line in \eqref{eq:GEB:sum} and apply an approach similar to the flat space argument above. 
The Dirichlet propagator may be represented in a momentum spectral representation
\begin{align}
    G_\textnormal{D}(\p;z,z')
    = 
    i \int \operatorname{d}\!m^2 
    \frac{ 
        f_{\textnormal{D}}(z, m^2) f_{\textnormal{D}}(z', m^2)
    }{
        \p^2 - m^2 + i \varepsilon
    }
\label{eq:GD_mom}
\end{align}
This representation  allows us to  apply the flat space cutting argument. A similar cut is presented in Ref.~\cite{Meltzer:2020qbr} for pure \AdS.
 
A hint for the correct combination of pseudoscalars is the bulk mixing between $\partial_\mu A^\mu$ and the field we denoted $\Theta$ in \eqref{eq:chiThetadef}. 
In \eqref{eq:SGFBulk} we fixed an $R_\xi$ gauge to cancel this mixing term. Further verification of this hint follows from recognizing that $\Theta$ has the same homogeneous equation of motion \eqref{eq:EOMTheta} as $A_\mu^\textnormal{L}$, \eqref{eq:EOM:and:B_op:L} for any dimension $d$ or bulk mass parameter $\alpha$. The propagators thus have the same form and the Dirichlet profiles are identical,
    \begin{align}
         f^\textnormal{L}_\textnormal{D}(z,m^2)=f^\Theta_\textnormal{D}(z,m^2) \ .
         \label{eq:f_D_rel}
    \end{align}
This is a necessary condition for the cancellations required by unitarity. 

Unitarity cuts are easy to implement in the momentum representation \eqref{eq:GD_mom}.  In  \tHF gauge, $\xi=1$, the unitarity argument for the Goldstone boson equivalence theorem follows that of flat space:
\renewcommand{\arraystretch}{2.5}
\begin{align}
       \vcenter{
        \hbox{\includegraphics[width=.25\textwidth]{{{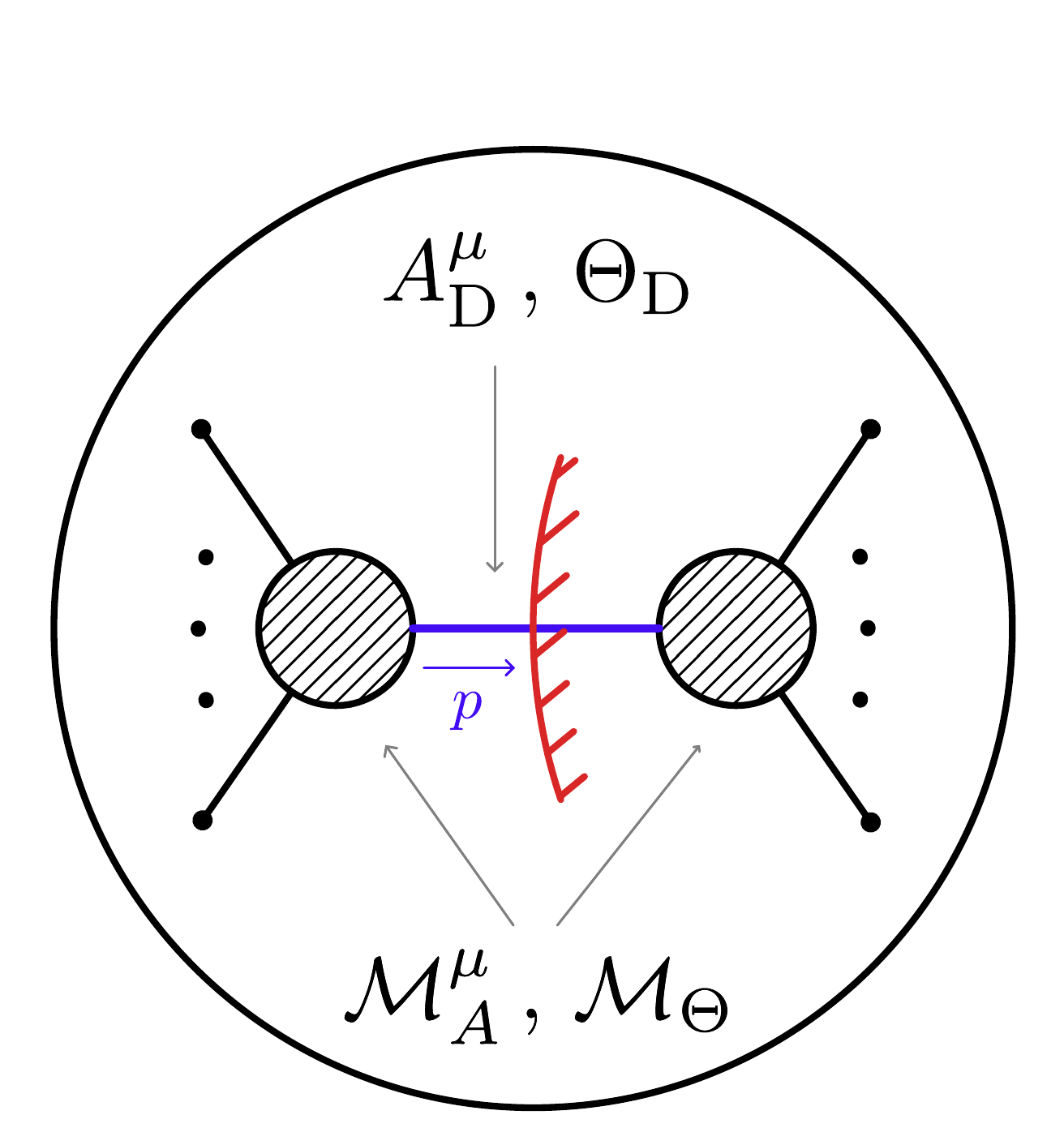}}}}
        }
        &
        &\Longrightarrow &
        &
        \begin{array}{ll}
        0
        = &
        \displaystyle
        -
        \left| 
            \frac{\p_\mu}{m}
            \int \Dz\, 
            f^\textnormal{L}_\textnormal{D}(z,m^2)
            \M_A^\mu(z, \p) 
        \right|^2 
        \\ &
        \displaystyle
        +
        \left| 
            \int 
            \Dz\, 
            f^\Theta_\textnormal{D}(z,m^2) 
            \M_\Theta (z, \p) 
        \right|^2 \ .
        \end{array}     
 \label{eq:GETAD:fig}
\end{align}
\renewcommand{\arraystretch}{1.1}
The cancellation from the diagram in \eqref{eq:GETAD:fig} is integrated over $dm^2$, but the result holds for any $k_\mu$ and is thus valid for each value of $m^2$.
We use the large-momentum limit of the longitudinal polarization \eqref{eq:GET:long:polz} to establish a bulk Goldstone boson equivalence theorem:
\begin{align}%
 \varepsilon^\textnormal{L}_\mu(\p) 
 \left.
    \int \Dz\, f^\textnormal{L}_\textnormal{D}(z,m^2)
    \M_A^\mu(z, \p)
 \right|_{\p^2 \gg m^2}
&\to 
    \int \Dz\, f^{\Theta}_\textnormal{D}(z,m^2)
    \M_\Theta (z, \p) \ .
 \label{eq:GETAD}
\end{align}
This equivalence theorem applies to diagrams whose external legs are Dirichlet modes. The crucial feature is the nontrivial matching of the $A^\mu_\textnormal{L}$ and $\Theta$ mode profiles, \eqref{eq:f_D_rel}. 
The result also holds for pure \AdS by sending  the \UV brane to infinity. In pure \AdS our Dirichlet modes become the \AdS normalizable modes and  the corresponding diagrams are  called  \AdS  transition amplitudes  \cite{Balasubramanian:1999ri}.

\subsection{Boundary Equivalence Theorem}
\label{sec:GET_boundary}

We now cut the boundary internal line in \eqref{eq:GEB:sum} and apply an approach similar to the flat space derivation. 
We expect that an equivalence theorem from this {cut} should involve the brane degrees of freedom.  Since $A_z$ has Dirichlet boundary conditions, the only scalar degree of freedom is the Goldstone boson $a$. 

{The cut lines are} the  brane-to-brane propagators of $A^\textnormal{L}_\mu$ \eqref{eq:b2b_G} and $a$ \eqref{eq:B_a:b2b_a}.  
{We observe that these brane-to-brane propagators have the structure of a $d$-dimensional free field propagator, 
$\tilde G_{\textnormal{UV},i}\equiv {\mathcal{B}^{-1}_i}$ dressed by insertions of the corresponding self-energy $\Sigma_{i}$, with $i=\textnormal{L},a$:}
\begin{align}
       \vcenter{
        \hbox{\includegraphics[width=.15\textwidth]{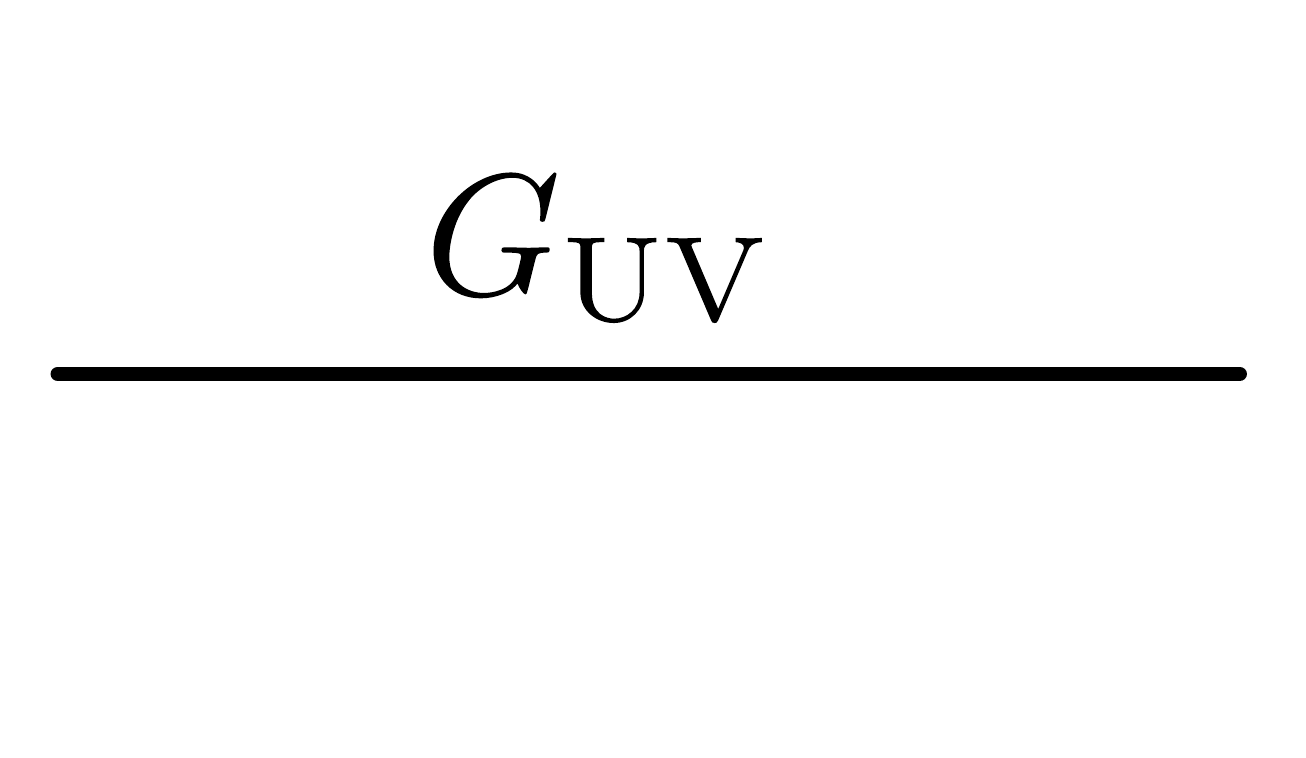}}
        }
        =
        \vcenter{
        \hbox{\includegraphics[width=.15\textwidth]{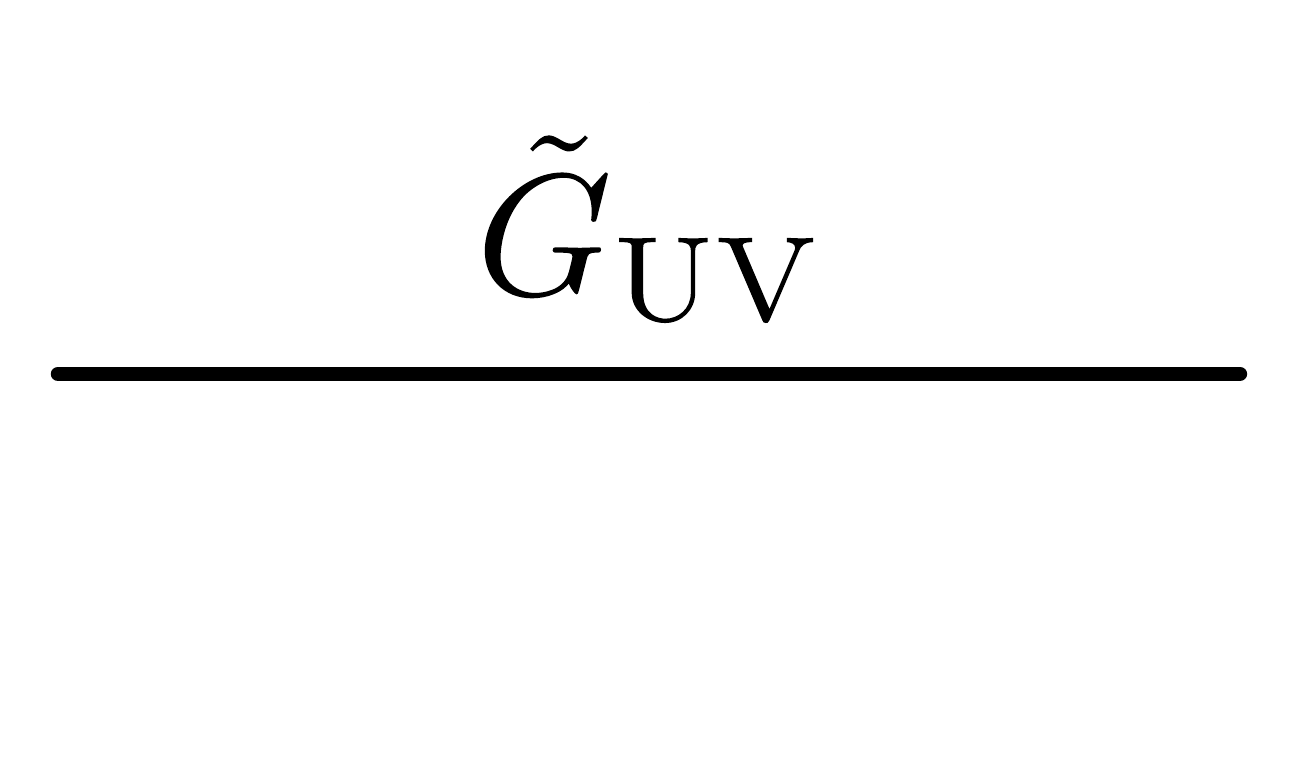}}
        }
        +
        \vcenter{
        \hbox{\includegraphics[width=.15\textwidth]{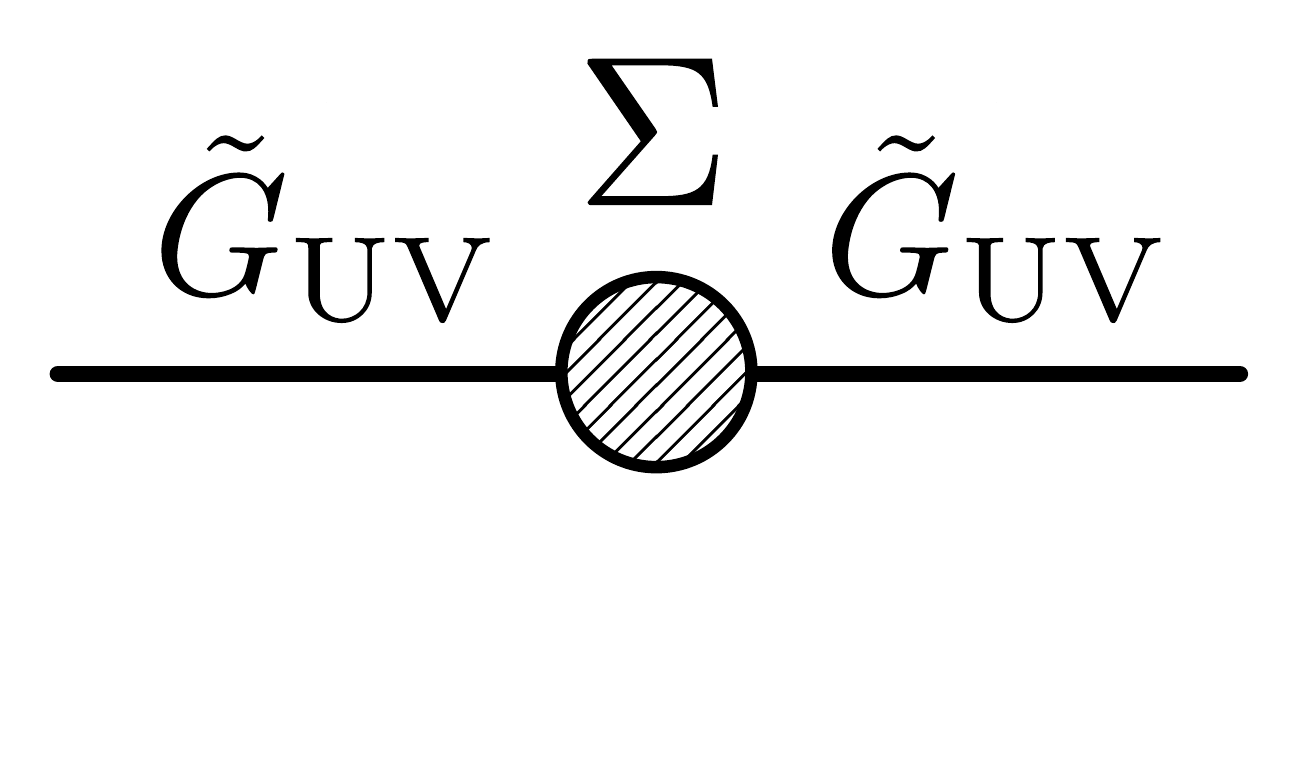}}
        }
        + 
        \cdots  .
        \label{eq:boundary:GET:geometric:series}
\end{align}
The  $\tilde G_\textnormal{UV}$ propagators are related by  
$\mathcal{B}_a
    =
    c_\textnormal{UV} 
    \xi_\textnormal{UV}\,
    {\g}^{-2}
    R^{d-3}
    \mathcal{B}_\textnormal{L}
$. 
Both feature an isolated pole at $%
    \p^2= c_\textnormal{UV} \xi_\textnormal{UV}v^2_\textnormal{UV}R^{d-3}
    $. 
We specialize to the case where the holographic self-energies are zero at this pole, $\Sigma_i|_\textnormal{pole}=0$.\,\footnote{In Section~\ref{sec:landscape} we show that this assumption is valid for $\alpha\leq \frac{d}{2}-1$.} When they are nonzero, the pole masses receive different corrections $\Sigma_a \neq \Sigma_\textnormal{L}$ that complicate the equivalence theorem. We leave this case for future investigation. 

Our remaining task is to identify an appropriate gauge. While the equivalence theorem is gauge-independent, we follow the Minkowski derivation which relies on the \tHF gauge where {the mass and normalization of the propagators are equal,} $G_\textnormal{T}=G_\textnormal{L}$. In our model, this gauge is realized with 
\begin{align}
    \xi &= 1
    &
    \xi_\textnormal{UV} &= \frac{ g^2 }{ \auv R^{d-3} }
    \,.
\end{align}
This choice implies that the propagators have the same pole mass,
\begin{align}
    \mathcal{B}_\textnormal{T} 
    =
    \mathcal{B}_\textnormal{L}  
    =
    \mathcal{B}_{a} 
    &= -\auv R \left(\p^2 - m_\mathcal{B}^2\right)\, 
    &
    m^2_\mathcal{B}
    &= 
    \frac{ c_\textnormal{UV} }{ \auv }
    \mA ^2 \,.
 \end{align}

The flat-space derivation instructs us to cut the internal $\tilde G_\textnormal{UV}$ lines. We obtain 
\renewcommand{\arraystretch}{2.5}
\begin{align}
       \vcenter{
        \hbox{\includegraphics[width=.25\textwidth]{{{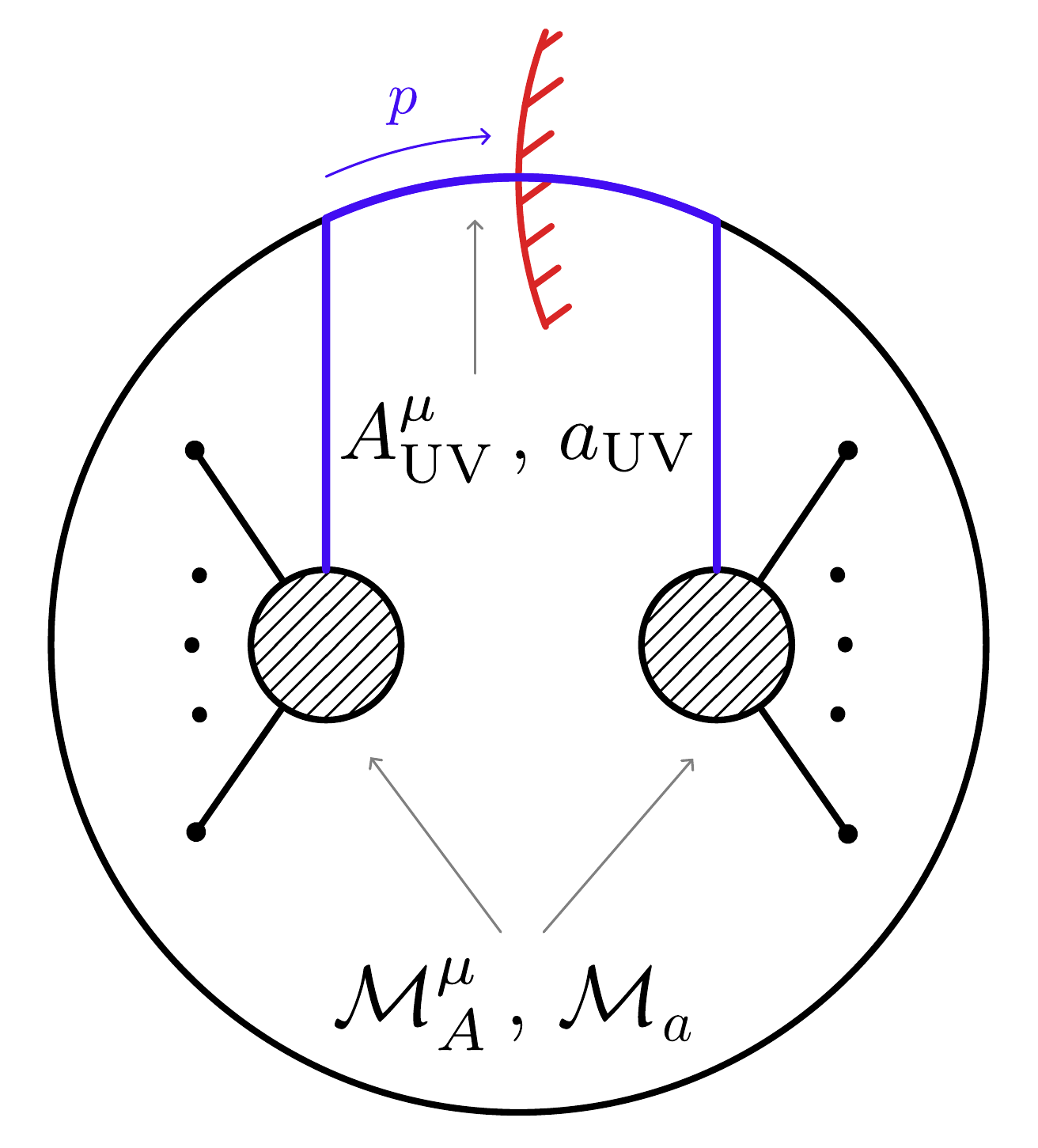}}}}
        }
        &
        &\Longrightarrow &
        &
        \begin{array}{ll}
        0 = &
        \displaystyle
        - \left|
            \frac{\p_\mu}{m_\mathcal{B}}
            \int \Dz\, 
            K(m_\mathcal{B}; z) 
            \M^\mu_{ A_\textnormal{UV}}(z, \p)
        \right|^2
        \\
        & \displaystyle
        +\left| 
            \int \Dz\, 
            K(m_\mathcal{B} ; z) 
            \M_{ a_\textnormal{UV}}(z,\p) 
        \right|^2 \ .
        \end{array}
        \label{eq:GET:boundary:equation}
\end{align}
\renewcommand{\arraystretch}{1.1}
Applying the large-momentum limit of the longitudinal polarization \eqref{eq:GET:long:polz} 
leads to a boundary version of the equivalence theorem:
\begin{align}
    \varepsilon^\textnormal{L}_\mu(\p) 
    \int \Dz\, 
        K( m_\mathcal{B} ; z ) 
        \M^\mu_{ A_\textnormal{UV} }(z,\p)
    \xrightarrow[\p^2 \gg m_\mathcal{B}^2]{} 
    \int \Dz\, 
        K( m_\mathcal{B} ; z )  
        \M_{ a_\textnormal{UV} }(z, \p)
    \ .
\label{eq:GET_flat}    
\end{align}
We see that this equivalence theorem applies to bulk amplitudes contracted with the boundary-to-bulk propagator evaluated at $m_\mathcal{B}$. The existence of a \tHF gauge for which the propagators of all three fields ($A_\textnormal{T}, A_\textnormal{L}$ and $a$) become equal is  a consistency check of our calculations.

\section{A Landscape of Broken U(1) in AdS}
\label{sec:landscape}

\begin{figure}[ht]
\centering
\includegraphics[width=\textwidth]{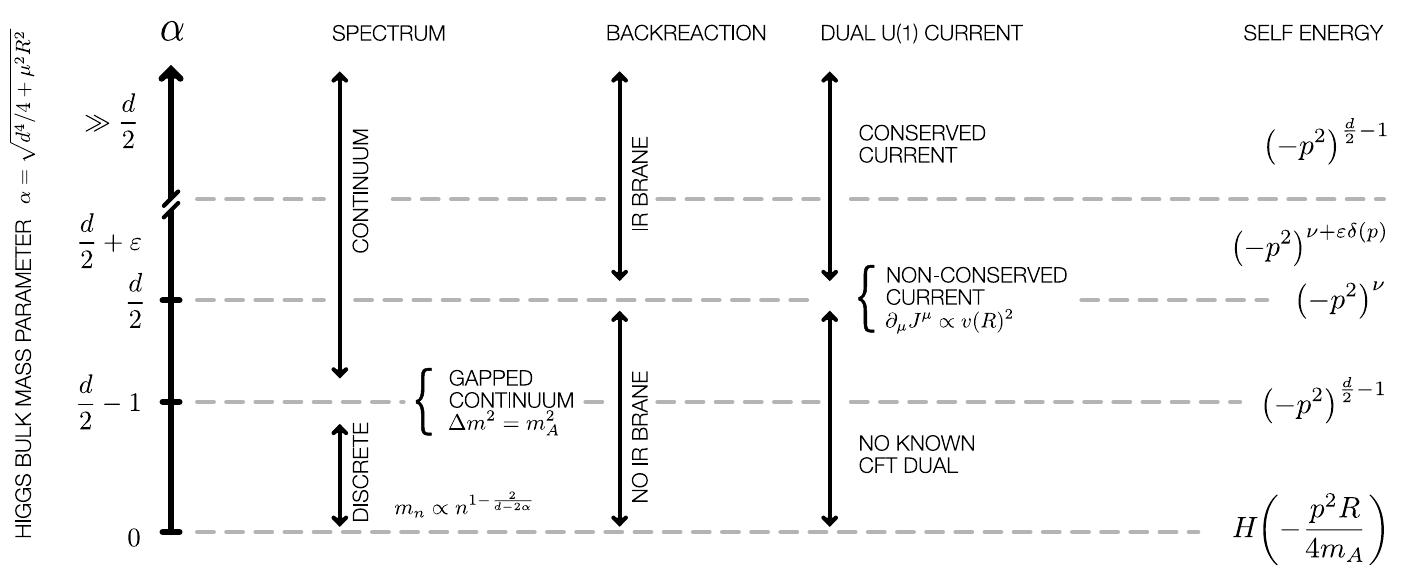}
\caption{%
    Summary of the vector holographic self energy and spectrum for a spontaneously broken \UU model in \AdSdpp as a function of the Higgs bulk mass parameter, $\alpha$.
    The last column shows the leading non-analytical term in the self energy. 
    \label{fig:summary}
}
\end{figure}

We turn to the derivation of the holographic self energy and spectrum  of the vector sector. In this section we write $A_\mu$ and $\Sigma$ for the transverse gauge field and its self energy.
The result for the longitudinal component is related by a rescaling of the Minkowski momentum, $p^\mu$, as we show in \eqref{eq:SigmaTL_rel}. 
We present results for non-negative Higgs bulk mass parameter, $\alpha \geq 0$. As discussed below \eqref{eq:vev:boundary}, the negative $\alpha$ case is physically identical. We summarize our results in Fig.~\ref{fig:summary}.

\subsection{WKB Approximation}
\label{sec:Schrodinger}

No closed form solution exists for the bulk gauge boson equation of motion~\eqref{eq:EOM:AT}. However, we may apply a \WKB analysis when $\left|pz\right| \gg 1$. 
To do this, we define a rescaled field ${\hat A_\mu =(z/R)^{(3-d)/2} A_\mu}$ that satisfies a Schr\"odinger-like equation of motion~\cite{Batell:2006pe},
\begin{align}
    \left[ \partial_z^2 - V(z)\right]\hat A_\mu &= -p^2 A_\mu
    &
    V(z)
    &
    =
    \frac{
        \left(d-3\right)
        \left(d-1\right)
    }{4z^2}
    + 
    g^2
    \left(\frac{R}{z}\right)^{\!2} v(z)^2 \ .
    \label{eq:schrodinger:pot}
\end{align} 
For  $\left|pz\right|\gg 1$, the approximate solutions for $\hat{A}_\mu$ are
\begin{align}%
    \exp\left(
        {\pm \int_z \Dz' \sqrt{V(z')-p^2} } 
        \right)   
    \underset{\left|pz\right| \gg 1}{\approx}
    \exp\left(
        \pm{\int_{z} \Dz' \sqrt{g^2\left(\frac{R}{{z'}}\right)^{\!2} v({z'})^2-p^2} } 
        \right) 
    \ .
    \label{eq:WKBsol} 
\end{align}
This provides an approximation for the spectrum of the gauge field in the Higgs background. The \vev behaves as $v\propto (z/R)^{\frac{d}{2}-\alpha}$, as described in \eqref{eq:vevsol}. 
Two regimes appear according to whether the $p^2$ or $v(z')^2$ term is dominant in \eqref{eq:WKBsol}:
\begin{itemize}
    \item \textit{$\alpha>\frac{d}{2}-1$ case.} The $p^2$ term is dominant so that the solutions asymptotically behave as if there is no Higgs background. The spectrum is therefore continuous.
    \item \textit{$\alpha\leq\frac{d}{2}-1$ case.}  The \vev term is dominant so that there is a turning point at $V(z_\textnormal{c})=p^2$ for timelike momentum. This indicates that the spectrum develops a mass gap. 
\end{itemize}

The cases are further distinguished by whether or not the \vev causes a significant gravitational backreaction in the bulk.  
For $\alpha \geq d/2$ the metric remains \AdS anywhere, while for $\alpha < d/2$ the metric is deformed at sufficiently large $z$ due to the growth of the \vev. 
In Section~\ref{sec:Stability} we model this backreaction with an infrared brane at a position $\zIR$ defined in \eqref{eq:zIR}. We analyze the effect of this aproximation on the vector spectrum in Section~\ref{sec:spectrum_backreaction}.

\subsection{Spectrum and Holographic Self-Energy} 
\label{sec:spectrum:and:self:energy}

We present approximate expressions for the transverse gauge field holographic self-energy $\Sigma$ for different values of the bulk mass parameter $\alpha$ relative to the dimension $d$.  For convenience, we define the low-momentum self-energy as the leading terms in a $|p \zUV| \ll 1$ expansion,
\begin{align}
     \Sigma(p) 
     \equiv 
     \Sigmalow(p)
     \left(1 + \mathcal O(|p\zUV|) \right) \ .
\end{align}
While it is common to choose coordinates where the \UV boundary is set to the radius of \AdS curvature $\zUV=R$, we write $\zUV$ for generality.
Here we present only the leading non-analytic term in $p^2$ since this encodes the continuum component of the spectrum. We present exact expressions and the leading analytic terms in Appendix~\ref{app:landscape:equations}.

\subsubsection*{Case: \texorpdfstring{$\alpha\gg \frac{d}{2}$}{Case: Alpha >> d/2}}

The Higgs \vev profile \eqref{eq:vevsol} decreases sharply with $z$. It thus affects the boundary condition at $\zUV$ but does not significantly affect the bulk profile of $A_\mu$.  The holographic self-energy is
\begin{align}
    \Sigmalow(p)
    &=
    \frac{2}{\zUV}
    \frac{ 
        \Gamma( 2 - \frac{d}{2} ) 
    }{
        \Gamma( \frac{d}{2} - 1 )
    }
    \left( \frac{ -p^2 \zUV^2 }{4} \right)^{\frac{d}{2}-1} 
    + (\text{analytic})
     \  ,
    \label{eq:Sigmalargealpha}
\end{align}
as in pure \AdS.

\subsubsection*{\texorpdfstring{Case: $\alpha = \frac{d}{2}\,\,(+\varepsilon)$}{Case: Alpha = d/2 (+epsilon)}}

The $\alpha=d/2$ case has an exact solution because the Higgs \vev term has the same scaling as a constant bulk mass for the gauge field. In fact, in this particular case the Higgs vev does not break the \AdS isometries.
The self energy is then
\begin{align}
    \Sigmalow(p)
     = 
    \frac{2}{\zUV}
    \frac{ \Gamma( 1 - \nu ) }{ \Gamma(\nu) } 
    \left( \frac{ -p^2 \zUV^2 }{ 4 } \right)^{\!\nu}     
    + (\text{analytic})
    \label{eq:Sigmaalpha:d/2}
\end{align}
with
\begin{align}
   \nu
    =
    \sqrt{
        \left(\frac{d}{2}-1\right)^{\!2}
        + m^2_A R^2
        }  
    \,. 
    \label{eq:nu}
\end{align}

For slightly larger values, $\alpha=d/2+\varepsilon$, the Higgs \vev may be understood as a slowly varying bulk mass for the gauge field. We perform a \WKB expansion with respect to $\varepsilon$ in Section~\ref{eq:CFTapprox}.\footnote{This a separate \WKB approximation from the one at large $|pz|$ that we describe in Section~\ref{sec:Schrodinger}.} To the best of our knowledge, this is the first analysis of the effect of a slowly varying bulk mass in \AdS. The holographic self energy is
\begin{align}
    \Sigmalow (p^2)
    &=
    \frac{2}{\zUV}
    \frac{ \Gamma( 1 - \nu ) }{\Gamma(\nu)}
    \left( \frac{-p^2 \zUV^2}{4} \right)^{\nu+\varepsilon\delta(p)} 
    + (\text{analytic})
     \ ,
\end{align}
where the function $\delta(p)$ varies slowly in $p$,
\begin{align}
    \delta(p) \approx  \frac{m_A^2 R^2}{2\nu} \log(p\zUV) \,.
\end{align}
The weak $p$-dependence of $\delta(p)$ reflects the soft breaking of the \AdS isometries by the Higgs \vev. The $\varepsilon=0$ case recovers the exact result of \eqref{eq:Sigmaalpha:d/2}. Correspondingly, conformal symmetry is only slightly broken in the holographic dual, see Section~\ref{eq:CFTapprox}. 

\subsubsection*{\texorpdfstring{Case: $\frac{d}{2}-1 < \alpha < \frac{d}{2} $}{Case: d/2 - 1 < alpha < d/2}}

Our \WKB analysis indicates that the spectrum is a continuum. Apart from this, there is no known approximation scheme to make quantitative statements in this regime.

\subsubsection*{\texorpdfstring{Case: $\alpha = \frac{d}{2}-1$}{Case: alpha = d/2}}

This case can be solved exactly. The low-energy limit of the self-energy is
\begin{align}
    \Sigmalow(p)
    & = 
    \frac{2}{\zUV}
    \frac{  \Gamma(2-\frac{d}{2})   }{  \Gamma(\frac{d}{2}-1)   }
    \left[
        \frac{(\mA ^2-p^2) \zUV^2}{4}
    \right]^{\frac{d}{2}-1} 
    + (\text{analytic})
     \label{eq:gappedselfenergyapprox}
     \ .
\end{align}
Here we assume that $d/2>1$ is not an integer and $\left|\mA ^2-p^2\right|\zUV^2 \ll 1$. The spectrum is a \textit{gapped continuum} that starts at $p^2=\mA ^2$. 

Equation \eqref{eq:gappedselfenergyapprox} does not hold for integer values of $d/2$. The Bessel functions must be expanded using the appropriate limits for integer order. Though a closed form expression is difficult to write, the source of non-analyticity is clear: the appropriate Bessel expansions contain logarithms. As an example, we present the case of $d=4$ which may be relevant to phenomenology. The holographic self-energy in the $|\mA ^2-p^2|\zUV^2\ll 1$ limit is
\begin{align}
    \Sigmalow (p)
    &
    =
    \frac{1}{2} \zUV 
    \left( p^2 - \mA ^2 \right)
    \left[ 
        \log\left( \frac{\zUV^2}{4} (p^2-\mA ^2) \right) 
    + 2\gamma_E
    - i \pi  \right]
    + (\text{analytic})
    \label{eq:gappedselfenergyintegerd}
\end{align}
where $\gamma_E$ is Euler's constant. As in the case where $d/2$ is not an integer, the gapped continuum begins at $p^2 = \mA ^2$.

\subsubsection*{\texorpdfstring{Case: $ 0 \leq \alpha < \frac{d}{2}-1$}{Case: alpha < d/2 - 1}}

The \vev dominates the potential at large $z$. Thus there is a \WKB turning point for timelike momentum and the spectrum contains discrete poles. These poles correspond to  the normalizable modes determined in the standard \WKB matching procedure. 
We can learn more about the discrete spectrum by using the \WKB approximation.
The spacing of the modes is given by 
\begin{align}%
    \Delta m_n^2
    =
    2 \pi 
    \left( \int^{z_\textnormal{c}(p)}_{\zUV} \Dz \frac{1}{\sqrt{V(z)-m_n^2}}  \right)^{-1}
    \label{eq:spacing}
\end{align} 
where $z_\textnormal{c}>\zUV$ corresponds to the turning point, $V(z_\textnormal{c})=p^2$.\,\footnote{A factor of $2$ seems to be missing in Eq.~(18) of \cite{Karch:2006pv}. } 
For $|p|\gg \mA $ the integral is dominated by the region near the turning point.
We find that the \WKB spectrum for $m_n \gg m_A$ scales like
\begin{align}%
    m_n 
    \propto 
    \mA      
    \left(
        \frac{n}{\mA  R}
    \right)^{1-\frac{2}{d-2\alpha}} \ .
    \label{eq:mnWKB:propto}
\end{align}
The exact result is given in \eqref{eq:mnWKB}. For $d=4$ and $\alpha=0$, the spectrum behaves as $m^2_n = 4 n {\mA }/{R}$ and exhibits a Regge behavior. 

\subsubsection*{\texorpdfstring{Case: $\alpha =0$}{Case: alpha = 0}}

The \WKB analysis for $\alpha < d/2 -1$ includes the case $\alpha = 0$. However, because the $\alpha=0$ case may be solved exactly, it is a useful check for \eqref{eq:mnWKB}. For $\mA  R \ll 1$, the non-analytic part of the holographic self-energy of $A^\mu$  is 
\begin{align}%
    \Sigmalow(p)
    &= 
    \frac{1}{2} \zUV\, p^2\, 
    H\!\left( -\frac{p^2 R}{4\mA  }\right) 
    + (\text{analytic}) \ 
\end{align}
where $H(x)$ is the analytically continued harmonic number. 
For timelike momentum, the harmonic number develops poles dictated by $H(1-r)=H(r) +\pi \textnormal{cotan}(\pi r)+$ finite. The poles occur at integer values of $r>1$. It follows that for $|p^2|R\gg \mA  $, the poles are $m^2_n = 4 n {\mA }/{R}$, exactly matching the \WKB result obtained in \eqref{eq:mnWKB:propto}.

\subsection{Infrared Backreaction and the Gauge Spectrum}
\label{sec:spectrum_backreaction}

\begin{figure}
    \centering
          \includegraphics[width=0.5\textwidth]{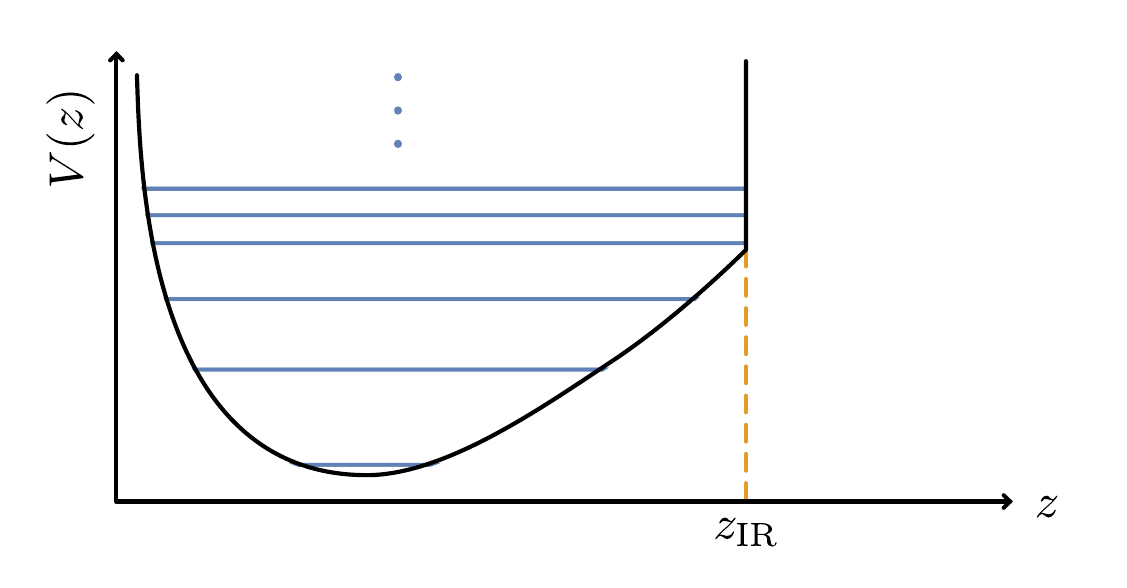}
    \caption{The effective Schr\"odinger potential for $\alpha<\frac{d}{2}-1$. 
    The presence of an \IR brane to model the Higgs backreaction only affects the upper part of the spectrum. 
    \label{fig:WKB}
    }
\end{figure}

For $\alpha<\frac{d}{2}$, the Higgs \vev grows with $z$ and induces a large metric backreaction. We model this effect by an \IR brane, see Section~\ref{sec:Stability}. Here we address how this approximation affects the gauge field spectrum.
Specifically, one may wonder to what extent a more detailed description of the backreaction is required to accurately determine the spectrum.  We distinguish the $\frac{d}{2}-1\leq\alpha<\frac{d}{2}$ and   $0\leq\alpha<\frac{d}{2}-1$ ranges.

For $\frac{d}{2}-1\leq\alpha<\frac{d}{2}$, the spectrum is  continuous if one ignores the \IR brane.  The presence of the \IR brane thus affects the spectrum   as a small discretization with $(\Delta m)_\textnormal{IR}\sim \pi/\zIR$ at any scale. We estimate the \IR brane position $\zIR$ in \eqref{eq:zIR}. We see that the spacing is parametrically small if $\vuv^2\ll M^{d-1}_*$, which is the natural mass hierarchy in the \EFT.  

Turning to $0\leq\alpha<\frac{d}{2}-1$, we show that  the bottom of the spectrum is solely determined by the shape of the Higgs \vev and is thus agnostic to the details of the large backreaction regime. In our \WKB analysis, the rescaled field $\hat A$ in \eqref{eq:schrodinger:pot} obeys Schr\"odinger-like potential $V(z)$ that we sketch in Fig.~\ref{fig:WKB}. Notice that the role of the \IR brane is only significant for modes where the \WKB turning point $z_\textnormal{c}$ reaches the \IR brane position $\zIR$. This amounts to a hard wall in $V(z)$ that divides the Schr\"odinger problem into two regimes:
\begin{itemize}
    \item For $z_\textnormal{c}(p)<\zIR$, the modes are \textit{not} sensitive to the \IR brane and are thus solely controlled by the Higgs background as dictated in \eqref{eq:mnWKB:propto}. In this regime, the gauge field spectrum is independent of how we describe the metric backreaction.
    \item In contrast, for $z_\textnormal{c}(p)>\zIR$, the gauge mode spacing is sensitive to the metric backreaction. Our \IR brane model of this regime predicts the typical spectrum for a slice of \AdS, $m_n\propto n$.
\end{itemize}
The  transition between the two regimes occurs at a mass $m_*$ given by $z_\textnormal{c}(m_*)=\zIR$. This translates into a condition on the mode number $n_*$ at which the transition occurs. We find  
\begin{align}%
    m_* 
    &\sim  
     \mA 
    \left( 
        \frac{ M^{\frac{d-1}{2}}_* }{\vuv}
    \right)^{ 1-\frac{2}{ d - 2\alpha } }
    &
    n_* & \sim
    \g R M_*^{\frac{d-1}{2}}
    \,,
    \label{eq:mstar:condIR} 
\end{align}
where we have used \eqref{eq:zIR} for $\zIR$ and \eqref{eq:mnWKB} to obtain $n_*$. These expressions are valid for any bulk mass in $\alpha\in[0,\frac{d}{2}-1)$. Interestingly, the critical mode number $n_*$ is independent on  $v_{\rm UV}$ and depends on $\alpha$ only via an $\mathcal O(1)$ prefactor. 

For $n<n_*$ the spectrum is controlled by the Higgs \vev profile and is not sensitive to the \IR brane. It is natural to have ${\g}^2M_*^{d-3}\gtrsim 1 $ and $RM_*\gg 1$ in the \EFT, which implies $n_*\gg 1$. The transition scale $m_*$ is analogously obtained by requiring that the mode spacing induced by the \IR brane is much smaller than the spacing induced by the Higgs \vev:
\begin{align}
    (\Delta m_n)_\textnormal{IR} \approx \frac{\pi}{\zIR} 
    & \ll 
    (\Delta m_n)_\textnormal{vev} \ ,
\end{align}
with $(\Delta m_n)_\textnormal{vev}$ derived above \eqref{eq:mnWKB}. This provides a check of the analysis.

\section{U(1) Breaking in the Holographic CFT }
\label{sec:CFT_Comments}

In this section we assume that the \AdSCFT conjecture applies to our \AdS Abelian Higgs model and  compute some properties of the resulting holographic \CFT. 
For $\alpha\geq \frac{d}{2}$ the Higgs \vev does not deform the \AdS metric. We can thus apply  \AdSCFT and identify the 
boundary effective action as the generating functional of a $d$-dimensional dual theory containing a conformal sector (\CFT).

We focus on two-point functions; these encode the information on conformal dimensions of the \CFT operators. We wish to understand the features encoded in the brane-to-brane propagator 
\begin{align}
    \langle
    A_\textnormal{UV}^\mu(p) A_\textnormal{UV}^\nu(-p)
    \rangle 
    &=
    - \frac{ i \eta^{\mu\nu} {\g}^2 }{ \mathcal{B}_G(p) + \Sigma(p) } 
    \ , 
    \label{eq:A02pt}
\end{align}
with $\alpha\geq \frac{d}{2}$  and  $pR \ll 1$ 
in terms of a $d$-dimensional \CFT model with broken \UU. \AdSCFT is valid in the  $pR \ll 1$ regime \cite{Aharony:1999ti}.

\subsection{Preliminary Observations}

The holographic dual theory is a $d$-dimensional \CFT coupled to boundary sources. The \CFT sector contains  a scalar primary $\cal O$  with conformal dimension 
\begin{align}
    \Delta_\mathcal{O}
    &= \frac{d}{2}+\alpha 
    \ ,
    \label{eq:Delta_O}    
\end{align}
and a \UU current $\cal J^\mu_\textnormal{CFT}$ with conformal dimension  $\Delta_\mathcal{J}$. Unitarity of the \CFT implies that a conserved current has $\Delta_\mathcal{J} = d-1 $ and a non-conserved current has  dimension
\begin{align}
    \Delta_\mathcal{J} 
    &= d-1 +\gamma_\mathcal{J} 
    \label{eq:Delta_J} 
    \ ,
\end{align} 
with $\gamma_\mathcal{J}>0$ \cite{Minwalla:1997ka}. 

In the presence of a \UV brane, the sources are dynamical fields in an \emph{elementary sector} {with} local operators. The couplings between \CFT and  elementary fields are irrelevant operators that deform the \CFT sector in the \UV. The elementary fields are identified with the boundary {degrees of freedom} in the holographic basis in Section~\ref{sec:hol_basis}:~\footnote{Here we define $\Phi(p;z) =\sqrt{R}\,\Phi_\textnormal{UV}(p) K(p;z) + \Phi_\textnormal{D}(p;z)$ in analogy to \eqref{eq:hol_basis}. The factor of $\sqrt{R}$ ensures that $\Phi_\textnormal{UV}$ has the dimension of a $d$-dimensional field.} 
\begin{align}
    \Phib &=\Phi_\textnormal{UV}
    &
    B^\mu &= A_\textnormal{UV}^\mu
    \ .
\end{align}

The Lagrangian of the $d$-dimensional dual theory has the general form
\begin{align}%
    \mathcal{L} 
    &= 
    \mathcal{L}_{\rm CFT}
    + \mathcal{L}_\textnormal{elem}[ \Phib, \Abmu ] 
    + \frac{ 
          b_\mathcal{J} 
        }{ 
          \Lambda^{ \Delta_\mathcal{J} - d+1 } 
        }
    \Abmu \mathcal{J}_\mu  
    + \frac{ 
          b_\mathcal{O}
        }{
          \Lambda^{\Delta_\mathcal{O} - \frac{d}{2} - 1 }
        } 
    \Phib \mathcal{O} 
    + 
    \text{h.c.} 
    + \cdots 
    \ ,
    \label{eq:LCFT}
\end{align}
where     $\Lambda \sim R^{-1}$ is the \CFT cutoff scale  and $b_\mathcal{J,O} \sim \mathcal O(1)$ are dimensionless coefficients. $\mathcal{L}_\textnormal{elem}$ encodes the elementary sector, including a \UU-breaking potential $V[\varphi]$, that we compute in Section~\ref{sec:effective:UV:potential}. The unspecified Lagrangian $\mathcal{L}_{\textnormal{CFT}}$ and the \CFT operators $\mathcal{J}_\mu  $, ${\mathcal O}$ depend on the underlying \CFT degrees of freedom whose dynamics is unspecified. 

The ellipses in \eqref{eq:LCFT} encode in principle further interactions between the $B^\mu$ field and the other elementary and CFT operators. These interactions may be needed in order to exactly reproduce the properties of the \AdS theory. 
For example, the $\Phib \mathcal O$ mixing term explicitly breaks the \UU symmetry. We can see this because the elementary field $\Phib$ that is dual to the bulk Higgs is \UU charged, whereas the \CFT operator $\cal O$ that it mixes with is uncharged.\,\footnote{
If $\mathcal O$ were charged, then it would be part of a conserved current. However, we cannot build such a current out of $\mathcal O$ unless $\mathcal O$ is a free field---for example, $\mathcal O\! \oset{\leftrightarrow}{\partial}_{\!\mu} \! \mathcal O^*$ is neither conserved nor has the correct dimension. }
For our goal in this section, the explicit terms in \eqref{eq:LCFT} are sufficient to understand the properties of the \CFT two-point functions.

\subsection{Holographic Effective Potential}
\label{sec:effective:UV:potential}

The \UU-breaking potential $V[\varphi]$ included in ${\mathcal L}_{\textnormal{elem}}$ is the zero-momentum part of the Higgs boundary effective action $\Gamma[\PhiUV]$.  At tree-level, we obtain this holographic effective potential by inserting the bulk \vev into the action $S_\textnormal{U(1)}$ in \eqref{eq:SD} and integrating by parts. By definition, the scalar wave operator vanishes when acting on $\langle\Phi\rangle$. This leaves a boundary action 
\begin{align}
    \Gamma_\textnormal{tree}
    =
    \int_\textnormal{UV} d^d x 
    \left[
    \Phi^\dagger \partial_z \Phi 
    +m_{\textnormal{UV}}^2 R \Phi^\dagger \Phi
    -
    \lambda_{\textnormal{UV}}R^{d-2}\left(\Phi^\dagger\Phi\right)^2
    \right] \ .
\end{align}
Using \eqref{eq:vevsol} to evaluate the $\partial_z$ term, and using the identification $\varphi=\PhiUV$, we obtain the effective potential 
\begin{align}
    V[\varphi]
    =
    -
    \frac{1}{R^2}
    \left( \frac{d}{2} - \alpha + m_{\textnormal{UV}}^2 R^2 \right) 
    \varphi^\dagger \varphi
    +
    \lambda_{ \textnormal{UV} } R^{d-2} 
    \left(\varphi^\dagger \varphi\right)^2
    \, .
    \label{eq:Veff}
\end{align}
We see that symmetry breaking occurs if ${d}/{2} - \alpha + m_{\textnormal{UV}}^2 R^2>0$, while no symmetry breaking occurs otherwise. This is precisely the combination that appears in $\vuv$ obtained in \eqref{eq:vevsol}.

\subsection{U(1) breaking}

\UU breaking in the dual theory is triggered by the effective
potential $V[\varphi]$ in $\mathcal{L}_\textnormal{elem}$, that induces a nonzero \vev for $\varphi$.  

\subsubsection*{Elementary sector}

The $\langle \varphi \rangle$ vev induces a mass for the elementary gauge boson, $m_{\!B}\propto g \langle\varphi\rangle$. 
Based on our Lagrangian \eqref{eq:LCFT} the two-point correlation function of $B_\mu$ is 
\begin{align}
    \langle
        B^\mu(p)\, B^\nu(-p)
    \rangle  
    &=
    \vcenter{
        \hbox{\includegraphics[height=.09\textwidth]{{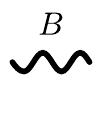}}}
        }
    +
    \vcenter{
        \hbox{\includegraphics[height=.09\textwidth]{{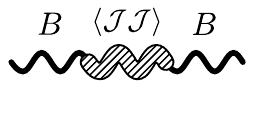}}}
        }
    +
    \vcenter{
        \hbox{\includegraphics[height=.09\textwidth]{{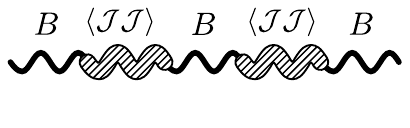}}}
        } 
    +
    \cdots
    \\
    &= 
    - 
    \left( 
        \eta_{\mu\nu} -
        \frac{p_\mu p_\nu}{p^2}
    \right) 
    \frac{
        i g^2
        }{
        Z_{\!B}\, p^2 -m^2_{\!B} + \kappa\,
        (-p^2)^{ \Delta_\mathcal{J} - \frac{d}{2} } 
        } 
    \,,
    \label{eq:B2pt}
\end{align}
where the $\kappa$ parameter is proportional to ${b^2_\mathcal{J}}{\Lambda^{-2\Delta_\mathcal{J} +d +2}}$. 
This is an exact result from dressing the free propagator of $B_\mu$ with bilinear insertions of $\langle \mathcal{J}^\mu \mathcal{J}^\nu \rangle$. 

\subsubsection*{CFT sector}

When $\langle\Phib\rangle\neq 0$, the $\Phib \mathcal{O}$ term is a deformation of the \CFT that breaks scale invariance unless  $\Delta_\mathcal{O} = d$. 
When the \CFT operator has dimension $\Delta_\mathcal{O}\gg d$, the deformation is highly irrelevant. Its effect is therefore negligible at energy scales much lower than $\Lambda$. 
For $\Delta_\mathcal{O} = d$ the deformation is exactly marginal  and thus does not break conformal invariance. 
In this latter case, we have  an exactly conformal sector with explicit $\UU$ symmetry breaking.

\subsection{Properties of the U(1) Current}

The \UU current is the sum of the elementary and the conformal sector currents: 
\begin{align}
    \mathcal{J}^\mu
    &=
    \mathcal{J}_\textnormal{elem}^\mu
    + \mathcal{J}_\textnormal{CFT}^\mu 
    \, . 
\end{align} 
The \UU symmetry is broken spontaneously in the elementary sector. As such, the elementary current is divergence-free, $\partial_\mu \mathcal{J}_\textnormal{elem}^\mu = 0 $. We examine the divergence of the composite sector current $\mathcal{J}_\textnormal{CFT}$ in the $\Delta_\mathcal{O}\gg d$ and $\Delta_\mathcal{O} = d$ cases. 

\subsubsection*{\texorpdfstring{$\Delta_\mathcal{O}\gg d$ Case}{Delta Large Case}}
\label{sec:DeltaLarge}

For large conformal dimension $\Delta_\mathcal{O}\gg d$, the $\langle\Phib\rangle \mathcal{O}$ deformation is highly irrelevant. Its effect at energies below the $\Lambda$ cutoff is thus negligible and the \CFT current is \textit{approximately conserved},
\begin{align}%
    \partial_\mu \mathcal{J}_\textnormal{CFT}^\mu
    &\approx 0 
    &
    \text{for}
    \quad
    \Delta_\mathcal{O}\gg d
    \;\;
    \text{and}
    \;\;
    p\ll \Lambda
    \,. 
    \label{eq:delJLargealpha}
\end{align}
Therefore the dimension of the \CFT current in this regime is $\Delta_\mathcal{J} = d-1$.

\subsubsection*{\texorpdfstring{$\Delta_\mathcal{O} = d$ Case}{Delta = d Case}}
\label{sec:Deltad}

When the conformal dimension of ${\cal O}$ is exactly $\Delta_\mathcal{O} = d$, the $\langle\Phib\rangle \mathcal{O}$ deformation is marginal, hence it does not break conformal symmetry.
The divergence of the \CFT current in the presence of this \UU-breaking   deformation is 
\begin{align}%
    \partial_\mu  \mathcal{J}_\textnormal{CFT}^\mu 
    & \sim  
    \frac{ 
        b_\mathcal{O}
        }{ 
        \Lambda^{ \frac{d}{2} - 1} 
        }
    \langle \Phib \rangle
    \mathcal{O}
    &
    \text{for}
    \quad
    \Delta_\mathcal{O} = d
    \;
    \text{and}
    \;
    p\ll \Lambda
     \,. 
     \label{eq:delJalphad}
\end{align}
Since the conformal symmetry is unbroken we can invoke properties of the conformal algebra and unitarity. We apply this to the states created by $\mathcal{J}_\textnormal{CFT}^\mu$ as follows~\cite{Minwalla:1997ka},
\begin{align}%
    |\, P_\mu \mathcal{J}^\mu|0\rangle \,|^2
    &=
    \langle 0|  
        \mathcal{J}^\nu K_\nu P_\mu \mathcal{J}^\mu
    | 0 \rangle
    =
    \gamma_\mathcal{J}
    \langle 0 |  
        \mathcal{J}_\mu \mathcal{J}^\mu
    | 0 \rangle
    \,
    \label{eq:PJ2}
\end{align}
where the generators of momentum $P$ and special conformal transformation $K$ are the raising and lowering operators of the \CFT.  Using $P_\mu = -i\partial_\mu$ and \eqref{eq:delJalphad},
we find that  that the anomalous dimension is proportional to the {squared modulus} of the symmetry breaking \vev, 
\begin{align}
    \gamma_\mathcal{J}
    \propto 
    \frac{ | b_\mathcal{O} |^2 }{ \Lambda^{d-2} }
    \left| \langle \varphi \rangle \right|^2  
    \,. 
    \label{eq:gam_CFT}
\end{align}

\subsection{Comparison to AdS}

The features obtained in the \CFT model above also emerge in \AdS. This supports our proposed model of a \CFT dual to our \AdS Abelian Higgs model.\footnote{Ref.~\cite{Karch:2023wui} performs a similar check  for a \CFT model with a double trace deformation.}
We compare the $\langle B^\mu \, B^\nu\rangle $ correlation function \eqref{eq:B2pt} to the corresponding \AdS brane-to-brane propagator $\langle A_\textnormal{UV}^\mu \, A_\textnormal{UV}^\nu\rangle $ \eqref{eq:A02pt}. The $\Delta_\mathcal{O} \gg d$ and $\Delta_\mathcal{O} = d$ cases correspond to $\alpha \gg d/2$ and $\alpha = d/2$, respectively.

\subsubsection*{\texorpdfstring{$\alpha\gg\frac{d}{2}$ Case}{alpha >> d/2 Case}}

The mass appearing in the \AdS boundary action  is $\sqrt{c_\textnormal{UV}}\,\mA $, which is  consistent with the scaling of $m_{\!B} \propto g\langle \Phib \rangle$. 
The holographic self-energy \eqref{eq:Sigmalargealpha} scales like $\Sigma(p)\propto (-p^2)^{\frac{d}{2}-1}$, which is consistent with the dimension of the \CFT current $\Delta_\mathcal{J}=d-1$ below \eqref{eq:delJLargealpha}.  
Finally, the fact that the elementary--\CFT mixing term $\varphi {\cal O}$ is a highly irrelevant operator is mapped in the \AdS picture to the sharp localization of the Higgs \vev toward the boundary so that it approximately does not influence the bulk equation of motion of the gauge boson.

\subsubsection*{\texorpdfstring{ $\alpha = \frac{d}{2}$ Case}{alpha = d/2 Case}}

The mass appearing in the \AdS boundary action is $\sqrt{c_\textnormal{UV} +({d-2})^{-1} }\,\mA $. This differs from the $\alpha \gg d/2$ case by a constant term from the self-energy, see \eqref{eq:Sigma:alpha:d:2}. 
This mass term is also consistent with the scaling of $m_{\!B} \propto g\langle \Phib \rangle$. 

The \AdS Higgs \vev introduces a quadratic term in $A_\mu$ that scales precisely as an \AdS bulk mass. As a result, the holographic self-energy scales as $\Sigma(p)\propto (-p^2)^{\nu}$, where $\nu$ is given by \eqref{eq:Sigmaalpha:d/2}.
For $\mA  R\ll 1$, we can isolate the \AdS computation of the \CFT anomalous dimension which we call $\gamma_{\textnormal{AdS}}$:
\begin{align}
    \nu
    &=
    \frac{d}{2} - 1 + \gamma_\textnormal{AdS}
    &
    \gamma_\textnormal{AdS} 
    &= 
    \frac{ (\mA  R)^2 }{ d-2 } 
    \,. 
     \label{eq:anom_current}
 \end{align}
Observe that $ \gamma_\textnormal{AdS}$ is proportional to the square of the Higgs \vev \cite{Nakayama:2013ssa}. 
This is consistent with the anomalous dimension  $\gamma_\mathcal{J}$ in the dual \CFT model, \eqref{eq:gam_CFT}. 
We can also see that this anomalous dimension is always  positive, which is consistent with the \CFT unitarity bound.

\section{Anomalous Dimension from a Near-AdS Background}
\label{eq:CFTapprox}

For a bulk mass parameter that is exactly $\alpha=\frac{d}{2}$, the classical background on which the gauge field propagates respects the isometries of \AdS. Bulk masses that are slightly perturbed from this value, $\alpha=\frac{d}{2}+\varepsilon$ with $\varepsilon \ll 1$, provide a background that \emph{approximately} retains the \AdS isometries. In that case a gauge field still experiences an effectively \AdS background over sufficiently small regions.

On the other hand, a brane-to-bulk propagator with momentum $p$ typically probes the region with $z\sim \frac{1}{p}$. The anomalous dimension of the \CFT $\UU$ current, $\gamma_\mathcal{J}$, arises from this specific region of the bulk.  

\begin{figure} 
    \centering
        \includegraphics[width=0.7\textwidth]{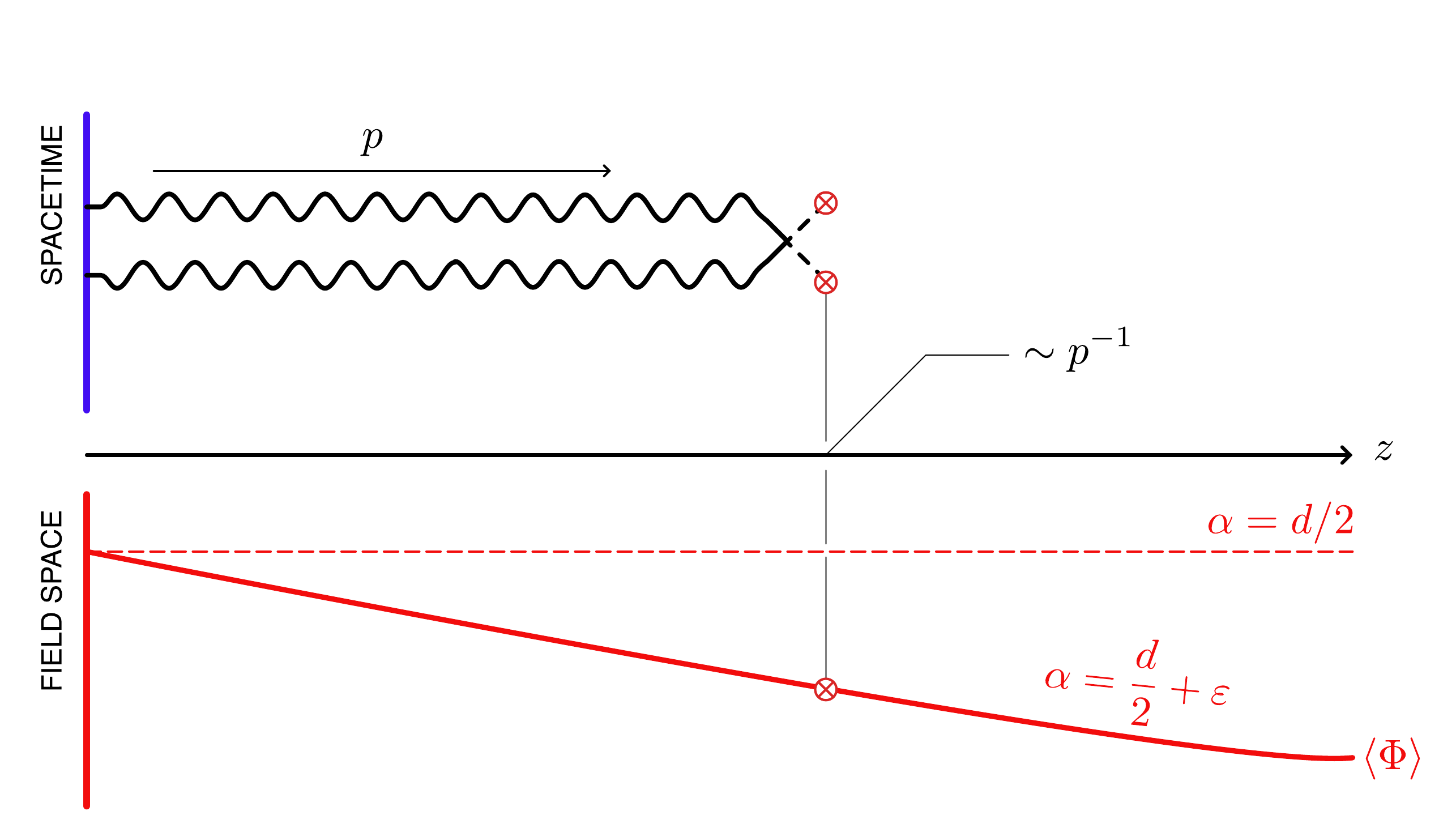}
    \caption{The slow variation of the Higgs \vev leads to a slowly running anomalous dimension for the \UU gauge field. \label{fig:WKBeps}}
\end{figure}

Combining the two above facts, we expect that if the background deviates slowly from \AdS as a function of $z$, then the anomalous dimension of the \CFT $\UU$ current should change slowly as a function of $p$. That is, $\gamma_\mathcal{J}$ should run as a function of the momentum scale with a $\beta$ function controlled by $\varepsilon$. This is pictured in Fig.~\ref{fig:WKBeps}.

One may also see this in the dual \CFT picture developed in Section~\ref{sec:CFT_Comments}. When $\alpha=\frac{d}{2}+\varepsilon$, the symmetry-breaking operator $\langle\Phib\rangle \mathcal O$ is approximately marginal so that the explicit breaking of the $\UU$ current changes slowly with the energy scale. This implies that the anomalous dimension $\gamma_\mathcal{J}$ runs, as in the \AdS picture. In this section we confirm these observations by explicitly computing the $\beta$ function of the anomalous dimension $\gamma_\mathcal{J}$ from the \AdS picture. 

The quantity we need to evaluate is the holographic self-energy of the transverse component of the gauge field, $\Sigma$. This self-energy is the logarithmic derivative of the boundary-to-bulk propagator $K$, \eqref{eq:holoselfenergy}. We drop the subscript `\acro{T}' with the understanding that we focus exclusively on transverse gauge fields. By construction, $K$ is given by the combination of solutions of the homogeneous equation of motion \eqref{eq:EOM:G} that is most regular as $z\to\infty$. For timelike momenta, this condition corresponds to an \emph{outgoing} boundary condition due to the  $+i\epsilon$ prescription. We assume timelike momenta, $p^2>0$, throughout the section.

\subsection{A near-AdS WKB approximation}
\label{sec:near:AdS:WKB}
To approach this problem, we use a \WKB approximation scheme where the expansion parameter is $\varepsilon$, the deviation in the bulk mass parameter $\alpha$ from its \AdS-preserving value $d/2$.\,\footnote{
This \WKB approximation differs from the one used to calculate the spectrum in Section~\ref{sec:Schrodinger} because there is a parametrically small deviation from a solution of Bessel functions.
} 
We apply this approximation to the transverse gauge boson, which we write simply as $A$ in this section. We transform the $A$ equation of motion \eqref{eq:B_op} into standard \WKB form by changing variables and rescaling the field, 
\begin{align}
    \y&= \log\!\left(\frac{z}{R}\right)
    &
    A(z) 
    &= \e^{ \dd \y } \psi(\y)
    &
    \dd &\equiv \frac{d}{2} - 1
    \, ,
    \label{eq:y:z:coords}
\end{align}
where $\y$ is a dimensionless position variable and $\dd$ is a function the spacetime dimension that appears often.
The equation of motion is 
\begin{align}
    \psi''(\y)+\k(\y)^2\psi(\y)
    &=0
    \quad
    &
    \quad
    \k(\y)^2
    &\equiv
    (R p)^2 \e^{2\y}
    -
    \dd^2
    -(R \mA )^2 
    \e^{-2\varepsilon \y}
    \ .
    \label{eq:EOM:psi:Q}
\end{align}
The crux of the \WKB approach is that, when the potential varies sufficiently slowly, the approximate solutions to \eqref{eq:EOM:psi:Q} are
\begin{align}
    \psi(\y) 
    &=
    \frac{1}{\sqrt{\k(\y)}} 
    \left[
    B^+
    \e^{+i\varphi(y)}
    +
    B^-
    \e^{-i\varphi(y)}
    \right]
    &
    \varphi(z)
    &\equiv
    \int_{\yt}^{y} \DY'\, Q(\y') 
\end{align}
as long as $\y$ satisfies the \WKB condition $|Q'(\y)|/Q(\y)^2 \ll 1$. 
The $\yt$ lower limit in the $\DY'$ integral is introduced for convenience. It is redundant with the normalization of the $B^\pm$ coefficients. These coefficients have mass dimension $[B^\pm]=d/2$.

\subsubsection*{Range of validity and turning point} 

For our potential, the \WKB condition is
\begin{align}
    \frac{\left|\k'(\y)\right|}{\k(\y)^2}
    =
    \left|
    \frac{
        (pR)^2 \, \e^{2\y} 
        + 
        \varepsilon \,
        (\mA  R)^2 \, 
        \e^{-2\varepsilon \y}
    }{
        \left[
            (pR)^2 \, \e^{2\y}
            -
            (d/2-1)^2
            -
            (\mA  R)^2 \, 
            \e^{-2\varepsilon \y}
        \right]^{3/2}
    }
    \right|
    \ll
    1
    \, .
    \label{eq:WKBlimit}
\end{align}
This condition is violated near a turning point where the potential vanishes,
\begin{align}
     \yt
     &=
     \log\!\left(
        \frac{\nu}{p R}
        \right)
        \left( 1+\mathcal O(\varepsilon) \right)
    &
   \nu
    &=
    \sqrt{
        \dd^2
        + \mA^2 R^2
        }  
     \, .
     \label{eq:WKB:turning:point}
\end{align}
The turning point separates the space into two disjoint domains where the \WKB condition is satisfied. The first is $\y\gg \yt$, corresponding to a large denominator in \eqref{eq:WKBlimit} because $p R \, \e^{\y} \gg 1$. The second is $\y \ll \yt$, corresponding to a small numerator. The latter case requires both $p R \, \e^{\y} \ll 1$ and $\varepsilon \ll 1$. Thus in the $\y\ll \yt$ region, the validity of the \WKB approximation is tied to the assumption that the bulk mass parameter $\alpha$ is close to $d/2$.

\subsubsection*{The WKB solutions} 

In the original $z$ coordinates, the general \WKB solutions in the asymptotic regions away from the turning point $z_\textnormal{t} = p/\nu$ are
\begin{align}
    A(z)
    &=
    \frac{z^\dd}{\sqrt{\k(z)}} 
    \begin{cases}
        B^+_<\,
        \e^{i\varphi(y(z))} 
        + 
        B^-_<\,
        \e^{-i\varphi(y(z))} 
        &
        z\ll z_\textnormal{t}
    \\
        B^+_>\,
        \e^{i\varphi(y(z))} 
        + 
        B^-_>\,
        \e^{-i\varphi(y(z))} 
        &
        z\gg z_\textnormal{t}
    \end{cases}
    &
  \quad\quad   y(z)=\log\left(\frac{z}{R}\right)
    \,.
    \, \label{eq:A_WKB:far}
\end{align}

\subsection{Matching in a Near-AdS Background}

To determine the $B$ coefficients of the \WKB solutions \eqref{eq:A_WKB:far}, we construct an intermediate region solution in the neighborhood of the turning point and match the asymptotic solutions to this intermediate region. 

\subsubsection*{Intermediate region Bessel solution} 

In the standard \WKB approach, one constructs an approximate solution around a turning point by assuming that $Q(\y)^2$ is linear in this region. Then exact solutions to the Schr\"odinger problem are Airy functions whose asymptotic expressions can be matched to the \WKB solutions that are valid away from the turning point. 
In our problem, using this linear $Q$ approximation would not utilize the fact that when $\varepsilon$ is zero, the potential has exact Bessel function solutions. Thus even in the $\varepsilon\to 0$ limit, the standard \WKB approach at finite order would introduce an irreducible error from forcing the expansion of the Bessel function into a basis of Airy functions. 

A more clever approach is to instead use the Bessel functions themselves as approximate intermediate-region solutions for $\varepsilon \neq 0$ . When $\varepsilon$ is small, the mass term in the equation of motion \eqref{eq:EOM:AT} picks up a weak scaling with the position, $m_A^2 (z/R)^{-2\varepsilon}$. Because $\varepsilon \ll 1$, we  can approximate this term with its value at the turning point, $z\approx\zt$. 
This amounts to an approximation of the potential
\begin{align}
    \k(z)^2
    &=
    p^2z^2
    -
    \nut^2
    + \mathcal O\!\left(\varepsilon \log\frac{z}{\zt} \right)
    \label{eq:Q:tilde}
\end{align}
where $\nut$ is $z$-independent but depends on momentum through $\zt$,
\begin{align}
    \nut
    \equiv
    \sqrt{
        \dd^2
        + \mA ^2 R^2 
        \left( \frac{\zt}{R} \right)^{-2\varepsilon}
    } 
    = \nu +\varepsilon\frac{m_A^2 R^2}{\nu}\log\left( \frac{pR}{\nu}\right)+{\cal O}(\varepsilon^2)
        \label{eq:nut}
\end{align}
In the region near the turning point,  the general solution  is then
\begin{align}
    \At(z)
    &=
    B_1\,
    z^\dd
    H^{(1)}_{\nut}(p z)
    +
    B_2\,
    z^\dd
    H^{(2)}_{\nut}(p z)
    &
    z \simeq z_t
    \ .
    \label{eq:A_WKB:zt}
\end{align}
This solution is valid in the intermediate region where the next-to-leading order term in \eqref{eq:Q:tilde} is negligible.

\subsubsection*{Asymptotic behavior}

We determine the coefficients of the asymptotic solutions \eqref{eq:A_WKB:far} by matching them to the near-turning point solution \eqref{eq:A_WKB:zt}.
To perform the matching, we construct  approximate asymptotic solutions using the same leading-order approximation \eqref{eq:Q:tilde} as the turning point solution, 
\begin{align}
    \All
    &=
     \frac{e^{\tfrac{-i\pi}{4}}}{\sqrt{\nut}}\, z^\dd
     \left[
            B^+_<\,
            \e^{-\nut}
            \left(
                \frac{pz}{2\nut}
            \right)^{-\nut}
            + 
            B^-_<\,
            \e^{+\nut}
            \left(
                \frac{pz}{2\nut}
            \right)^{+\nut}
        + 
        \right]
        \left[1+\mathcal O\! \left( \frac{pz}{2\nut} \right)\right]
    &
    z&\ll \zt
    \label{eq:WKB:All}
    \\
    \Agg
    &=
    \frac{z^\dd}{\sqrt{pz}}
    \left[        
    B^+_>\,
    \e^{ + i \left( pz - \tfrac{\pi\nut}{2} \right) } 
    + 
    B^-_>\,
    \e^{ - i \left( pz - \tfrac{\pi\nut}{2} \right) } 
     \right]
    \left[1+\mathcal O\! \left( \frac{2\nut}{pz} \right)\right]        
    &
    z&\gg \zt
    \, .
    \label{eq:WKB:Agg}
\end{align}

\subsubsection*{Matching for large $z$} 

We impose that plane waves are purely outgoing at the Poincaré horizon, $z\to\infty$.  With the standard $p^2 = |p^2|+i\epsilon$  prescription for timelike momenta, this outgoing boundary condition is equivalent to regularity for large $z$. The boundary condition thus imposes that the solution proportional to $\e^{ - ipz}$ in \eqref{eq:WKB:Agg} has zero coefficient, $B^-_> = 0$.
 
To match this outgoing solution to the intermediate region solution, we expand the Hankel functions in \eqref{eq:A_WKB:zt} with respect to their large argument limit:
\begin{align}
    H_{\nut}^{(1,2)}(pz) 
    & \xrightarrow[pz\to\infty]{}
    \sqrt{ \frac{2}{ \pi pz } } 
    \e^{ \pm i \left(p z -\tfrac{{\nut} \pi}{2} - \tfrac{\pi}{4} \right)} 
    \ .
\end{align}
The condition $B^-_> = 0$ thus matches to $B_2=0$ on the intermediate region solution. The $B_1$ coefficient is proportional to $B^+_>$. 

\subsubsection*{Matching for small $z$} 

The small argument limit of the remaining Hankel function in intermediate region is,
\begin{align}
    H^{(1)}_{\nu_t} (p z)
    & \xrightarrow[pz\to 0]{}
    - \frac{i}{\pi} 
    \left[ \Gamma(\nut) 
        \left( \frac{p z}{2} \right)^{-\nut}
        + \Gamma(-\nut) \e^{-i\pi\nut} \left( \frac{p z}{2} \right)^{\nut}    
    \right]
    \label{eq:Hankel1:asym}
\end{align}
The $pz$ scaling in each of the two terms exactly matches those of the two terms in the $z\ll \zt$ limit of $A=\All$ in \eqref{eq:WKB:All}. We find that the relative coefficients of the two terms is
\begin{align}
    \frac{B_<^-}{B_<^+}
    &= \e^{-i\pi \nut -2\nut }
    \nut^{2\nut } 
    \frac{ \Gamma(  - \nut )}{\Gamma(\nut)}
    \label{eq:ratioWKBbessel}
    \, .
\end{align}

\subsubsection*{Full WKB Solution}

The complete \WKB solution \eqref{eq:A_WKB:far} in the $z\ll \zt$ limit uses the relative coefficients from the turning point matching, \eqref{eq:ratioWKBbessel},
\begin{align}
    A(z)
    &\propto 
    \frac{z^\dd}{\sqrt{\k(z)}} 
    \left(
        \e^{i\varphi(z)} 
        + 
        \frac{B_<^-}{B_<^+}
        \e^{-i\varphi(z)}
    \right)
    \ .
    \label{eq:WKB:full:A:solution:implicit}
\end{align}
We calculate the phase $\varphi$ to first order in $\varepsilon$,

\begin{align}
    -i \varphi(z)
    &= 
    \nu + \nu\log\left(\frac{p z}{2\nu}\right)
    +
    \varepsilon \frac{\mA^2 R^2}{2\nu}
    \left[
    \log^2\left(\frac{p z}{2\nu}\right) 
    -\log^2\left(\frac{ z}{R}\right) 
    -\frac{\pi^2}{12} 
    \right]
      +
    \mathcal O\!\left(\varepsilon^2,\, p^2z^2\right)  
  \label{eq:phismall}
\end{align}

\subsection{Near-AdS Holographic Self-energy }

The holographic self-energy is
\begin{align}
    \Sigma(p) = - \left.\partial_z\log A\right|_{\zUV} \ , \label{eq:Sigma_def}
\end{align}
where $A$ is the regular solution of the equation of motion. Since $\Sigma$ is a logarithmic derivative evaluated at the boundary, we only need the solution in the $z\ll \zt$ regime. 
Moreover, the overall normalization of $A(z)$ cancels, thus we only require the ratio of coefficients, $B^-_</B^+_<$ in \eqref{eq:ratioWKBbessel}. This ratio depends on momentum through $\nut$, as shown in \eqref{eq:nut}. 
By expanding the logarithm of \eqref{eq:ratioWKBbessel} at $\mathcal O(\varepsilon^2)$, we obtain
\begin{align}
    \frac{B_<^-}{B_<^+}
    = \e^{-i\pi \nu }
    \nu^{2\nu } 
    \frac{ \Gamma(  - \nu )}{\Gamma(\nu)} 
    \left( \frac{p R}{\nu} \right)^{
        \frac{ \varepsilon m_A^2 R^2 }{ \nu }
        \left( -i \pi + 2 \log\nu - \Psi_\nu - \Psi_{-\nu} \right)
        } 
    \label{eq:ratio}
\end{align}
where $\Psi_\nu={\Gamma'(\nu)}/{\Gamma(\nu)} $ is the digamma function.

$A(z)$ contains a factor of $Q^{-1/2}$ in \eqref{eq:WKB:full:A:solution:implicit}. This factor only contributes to the self energy at order $\mathcal O(p^2\zUV^2)$.
The phases in \eqref{eq:ratio} combine with the logarithms in the self-energy at leading and next-to-leading order.\footnote{%
    At next-to-leading order this is obtained from the identity 
    \begin{align*}
    \log^2\!\left( \frac{p z}{2\nu} \right)
    - i \pi \log\left( \frac{ p z }{ 2 \nu } \right) 
    &= 
    \frac{1}{4}
    \log^2\left( \frac{ - p^2 z^2 }{ 4 \nu^2 } \right)
    - \frac{ 3 \pi^2 }{ 4 }
\end{align*}
 using the $p^2=|p|^2+i\epsilon$ prescription to pick the branch cuts.
} 
The resulting expression for the self energy is
\begin{align}
    \Sigma(p)
    &=
    \frac{\nu-h}{\zUV}
    + 
    \frac{2}{\zUV}
    \frac{\Gamma(1-\nu)}{\Gamma(\nu)}
    \left(-\frac{p^2 \zUV^2}{4}\right)^{\!\nu+\varepsilon\delta(p)} 
    +\mathcal O(\varepsilon^2,\, p^2\zUV^2)
    \label{eq:selfenergy:WKB}
    \ ,
\end{align}
where the correction to the power of the non-analytic term is
\begin{align}
    \delta(p) 
    = 
    \frac{  m_A^2 \zUV^2 }{ 2 \nu } 
    \left(
        \log\left(\sqrt{-p^2} \zUV\right)
        +\mathcal O(1) 
    \right)
    \ . 
    \label{eq:delta}
\end{align}
We neglected the $\mathcal O(\varepsilon)$ constants in \eqref{eq:delta} because
the logarithm dominates for $z \ll \zt$. The fact that $\log(-p^2)$ appears in $\delta(p)$ ensures that we can consistently continue the result to spacelike momenta $p^2<0$. In that case the self-energy becomes real at the order of approximation we consider. This is a consistency check of our \WKB calculation. 

Our result \eqref{eq:selfenergy:WKB}  matches  the self-energy obtained in the  \AdS limit \eqref{eq:Sigma:alpha:d:2} within the order of approximation considered, 
except that the power of the non-analytical term has logarithmic dependence on momentum. This logarithmic dependence  reproduces the intuition of a logarithmic running of the anomalous dimension. 

The anomalous dimension $\gamma_\mathcal{J}$ in the pure \AdS case is identifed in \eqref{eq:anom_current}. The $\delta(p)$ corrections encodes how this anomalous dimension {runs} in the near-\AdS case due to the near-marginal deformation of the \CFT. The $\beta$ function  coefficient of $\gamma_{\cal J}$ is extracted from $\delta(p)$ by taking the log derivative in $p$ or $\zUV$:  
\begin{align}
    \beta_{\gamma_\mathcal{J}} 
    &= \varepsilon \frac{ d }{ d \log \zUV} \delta(p ) 
     = \varepsilon \frac{\mA ^2 R^2}{2\nu} \ .
\end{align}

\section{Summary}

Our \AdSdpp Abelian Higgs model contains a {bulk} \UU gauge field and a bulk Higgs.  A potential on the \UV brane induces a Higgs \vev that extends into the bulk. 
Using $R_\xi$ gauges on the bulk and brane, we derive the bulk propagators and boundary effective actions of the gauge field and pseudoscalar sectors. In doing so, we highlight subtle points with gauge fixing, boundary conditions, and the linear combination of pseudoscalars that mix with the vector. We then present a variety of features  of the theory:
\begin{itemize}   

    \item  In Section~\ref{sec:GET} we identify two distinct \AdS Goldstone equivalence theorems by using a holographic decomposition of the fields. The bulk equivalence theorem involves a specific combination of the Goldstone and $A_z$. The boundary equivalence theorem involves isolated brane-localized modes.  

    \item The bulk Higgs mass parameter,\footnote{This is related to the Higgs bulk mass by $\mu^2=(\alpha^2-\frac{d^2}{4})R^{-2}$.} $\alpha$, controls the gauge spectrum. In Section~\ref{sec:landscape}, we show that the spectrum may be continuous ($\alpha>\frac{d}{2}-1$), continuous with a mass gap ($\alpha=\frac{d}{2}-1$), or discrete ($0\leq\alpha<\frac{d}{2}-1$) We summarize this in Fig.\,\ref{fig:summary}. We compute the discrete spectrum for any $\alpha$ and  compute the holographic self-energies for certain ranges of $\alpha$. The spectrum contains a pole corresponding to a $d$-dimensional gauge mode which is massive due to symmetry breaking. 

    \item The bulk \vev induces a significant backreaction of the metric when $\alpha<\frac{d}{2}$. We model the effect of this backreaction with an \IR brane. 
    For $\frac{d}{2}-1\leq \alpha <\frac{d}{2}$, the brane induces a discretization with negligible spacing.
    For $\alpha < \frac{d}{2} - 1$, we show the low energy spectrum only depends on the background near the \UV brane and is thus insensitive to the details of the backreaction.

    \item We turn to the \CFT dual of the theory in Section~\ref{sec:CFT_Comments}.
    For $\alpha\gg \frac{d}{2}$, the \CFT sector of the holographic dual theory is not affected by the \UU breaking since the \CFT current is conserved. For $\alpha =\frac{d}{2}$, the dual \CFT sector contains an exactly marginal operator that explicitly breaks the \UU symmetry of the \CFT sector. The \UU~\CFT current is thus not conserved and we show that its anomalous dimension is proportional to the square of the \UU-breaking \vev. This result parametrically matches the anomalous dimension computed in the \AdS theory.  

    \item In Section~\ref{eq:CFTapprox} we study the theory with bulk mass parameters near $\alpha =\frac{d}{2}$. In this regime the \AdS isometries are approximately preserved by the Higgs vev. We introduce a \WKB method to solve the field equations in a near-\AdS background. The resulting boundary action describes an approximately conformal \UU current whose dimension runs logarithmically with the energy scale. On the \CFT side, this encodes the fact that the \UU breaking deformation is nearly-marginal.  

\end{itemize}

\section*{Acknowledgments}

Kuntal Pal contributed to early aspects of this study. We thank
Csaba Cs\'aki,
Lexi Costantino,
Liam Fitzpatrick,
Gero von Gersdorff,
Eugenio Megias,
David Meltzer,
Maxim Perelstein,
Mariano Quiros 
and
Brian Shuve
for insightful discussions.
\acro{PT} thanks 
    the Aspen Center for Physics (\acro{NSF} grant \#1066293), 
    the Kavli Institute for Theoretical Physics (\acro{NSF\,PHY}-1748958 and \acro{PHY}-2309135),
    the 2022 Pollica Summer Workshop on Dark Matter,\footnote{The Pollica Summer Workshops are by the Regione Campania, Università degli Studi di Salerno, Università degli Studi di Napoli ``Federico II'', i dipartimenti di Fisica ``Ettore Pancini'' and ``E.~R.~Caianiello'', and Istituto Nazionale di Fisica Nucleare.} 
    the \acro{SAIFR}/Principia Workshop on the Nature of Dark Matter,
    and the Center for Theoretical Underground Physics and Related Areas\footnote{Center for Theoretical Underground Physics and Related Areas (CETUP*), The Institute for Underground Science at Sanford Underground Research Facility (SURF), and the South Dakota Science and Technology Authority} (\acro{CETUP}$^{*}$)
    for their hospitality during a period where part of this work was completed. 
\acro{PT} is supported by an \acro{NSF CAREER} award (\#2045333).

\appendix
\part*{Appendices}
\addtocontents{toc}{\protect\setcounter{tocdepth}{1}}

\section{An Exact Gravitational Backreaction}
\label{se:softwall}

Throughout this paper we approximate the gravitational backreaction occurring in the \acro{IR} for bulk mass parameter $\alpha<\frac{d}{2}$ with an effective infrared brane whose location is estimated in \eqref{eq:zIR}. We check the validity of this approximation by comparing it to an exact solution of the Higgs--gravity system that creates a so-called soft wall. These exact solutions are conveniently derived using the superpotential formalism, see e.g. \cite{DeWolfe:1999cp,Papadimitriou:2007sj,Megias:2018sxv}.\footnote{
We thank Eugenio Megias for insightful explanations regarding these points. 
}
It is convenient to work with the proper coordinate $r$ where the warped metric ansatz is
\begin{align}
    \D{s}^2
        = e^{-2A(r)} \eta_{\mu \nu}\D{x}^\mu \D{x}^\nu-\D{r}^2
    \label{eq:metric_gen } \, .
\end{align}
The relation to the conformal coordinate $z$ is $ e^{A(r)}dr =dz$. In the \AdS limit $e^{\frac{r}{R}}=z/R$.

The simple bulk potential used in this text, $V[\Phi]=\mu^2 \Phi^\dagger \Phi$ with $\mu^2 R^2 \equiv \alpha^2 - \frac{d^2}{4}$, does not lend itself to a simple solution for the $\Phi$ vacuum expectation value when the gravitational backreaction cannot be ignored, $\alpha < d/2$. One must iteratively solve for the vev $\langle \Phi \rangle = v(r)$, its gravitational backreaction, and then revise $v(r)$ due to this backreaction correction, and so forth. Instead, one may consider a slightly different  potential motivated by the superpotential formalism,\,\footnote{
The superpotential formalism for any dimension can be found in \cite{Fichet:2023xbu}. The superpotential used here is 
$W=\frac{1}{R}(d-1)(d-2)M_*^{d-1}+\frac{1}{R}(d-2)(\frac{d}{2}-\alpha)|\Phi|^2$. 
}
\begin{align}
    V[\Phi]
    = 
    \frac{\alpha^2 - \frac{d^2}{4}}{R^2}
    \Phi^\dagger \Phi 
    - \frac{ d\left(\frac{d}{2} - \alpha\right)^2 }{ 2(d-1) R^2 M_*^{d-1} } 
    (\Phi^\dagger \Phi)^2
    \,.
    \label{eq:super:potential:inspired:V}
\end{align}

The exact solutions for the Higgs vev and the metric  are~\cite{Megias:2018sxv,Megias:2020vek}
\begin{align}
v(r)&=  v_\textnormal{UV}
        e^{\left( \frac{d}{2} - \alpha \right)
        \frac{r}{R}}
&&
&A( r)&= \frac{r}{R} + \frac{v_{\rm UV}^2}{4(d-1) M_*^{d-1}}
        e^{\left(d-2\alpha\right)\frac{r}{R}} 
        \,. \label{eq:softwall}
\end{align}
The backreaction only appears as a nonlinear modification to the exponent of the warp factor, $A(r)$. The exact Higgs vev profile precisely matches what we find from the infrared brane approximation, \eqref{eq:vevsol}.

The spacetime is asymptotically \AdS at small $r$ i.e.\ small $z$. 
The backreaction is significant when the \vev term in \eqref{eq:softwall} is of the same order of magnitude as the \AdS term, $r/R$.
In conformal coordinates this condition is
\begin{align}
    z^{d-2\alpha}_\textnormal{IR}
    \sim R^{ d - 2\alpha } 
    \frac{4(d-1) M_*^3 }{ v_\textnormal{UV}^2 }
    \log\left(\frac{z_\textnormal{IR}}{R}\right) \ .
    \label{eq:significant:backreaction:condition}
\end{align}
Observe that \eqref{eq:significant:backreaction:condition} has the same parametric form as the general estimate given in \eqref{eq:zIR}. This provides a nontrivial check of our approximate approach to the backreaction.

We can readily compute the Schrödinger-like potential following the formalism of Sections~\ref{sec:Schrodinger} and \ref{sec:spectrum_backreaction} using the exact soft wall metric \eqref{eq:softwall}. For $z_\textnormal{c}(p) < z_\textnormal{IR}$, the the effect of the backreaction is negligible and the general conclusion applies: the bottom part of the spectrum is independent of the \IR backreaction.

\section{Evaluation of the Gauge Boson Self-Energy}
\label{app:landscape:equations}

\begin{table}[!ht]
    \renewcommand{\arraystretch}{1.7} 
    \setlength{\tabcolsep}{20pt} 
    \begin{tabular}{ @{} rll @{} } \toprule 
        Bulk mass 
        &
        Regular Solution $(z\to\infty)$
        &
        Second Solution 
        \\
        \midrule
        $\alpha \gg \frac{d}{2}$
        & $z^{\dd} K_{\dd}(\sqrt{-p^2z^2})$
        & $z^{\dd} I_{\dd}(\sqrt{-p^2z^2})$
        \\
        $\alpha = \frac{d}{2}$
        & $z^{\dd} K_{\nu}(\sqrt{-p^2z^2})$
        & $z^{\dd} I_{\nu}(\sqrt{-p^2z^2})$
        \\
        $\alpha = \frac{d}{2}-1$
        & $z^{\dd} K_{\dd}(\sqrt{\mA^2z^2 - p^2z^2})$
        & $z^{\dd} I_{\dd}(\sqrt{\mA^2z^2 - p^2z^2})$
        \\
        $\alpha = 0 $
        & $\displaystyle \exp\!\left(\frac{\mA }{2R}z^2\right) L\!\left( \frac{-p^2 R}{4\vuv}, 0, \frac{\mA  z^2}{R} \right)$
        & $\displaystyle \exp\!\left(\frac{\mA }{2R}z^2\right) U\!\left( \frac{-p^2 R}{4\vuv}, 0, \frac{\mA  z^2}{R} \right)$
        \\
        \bottomrule
    \end{tabular}
    \caption{Solutions to the homogeneous bulk equation of motion for different values of the Higgs bulk mass parameter $\alpha^2 = \tfrac{d^2}{4}+\mu^2R^2$. We use the solution that is regular at $z\to\infty$ to calculate the holographic self energy. We write $h = \tfrac{d}{2}-1$ and $\nu^2 = h^2 + \mA^2 R^2$. We examine the case where $\alpha \approx \tfrac{d}{2}$ separately in Section~\ref{eq:CFTapprox}.  For the case $\alpha = 0$, $U$ is the confluent hypergeometric function and $L$ is the Laguerre polynomial. 
    }
\label{tab:solutions}
\renewcommand{\arraystretch}{1.1} 
\end{table} 

We complement Section~\ref{sec:spectrum:and:self:energy} with additional details for the computation of the holographic self energy and the  spectrum. Table~\ref{tab:solutions} summarizes the solutions to the homogeneous bulk equation of motion. The solution that is regular for spacelike momentum at $z\to\infty$ corresponds to an outgoing plane wave for timelike momentum as dictated by the $i\epsilon$ prescription.  The holographic self energy \eqref{eq:holoselfenergy} is a logarithmic derivative of this solution. For convenience, we define
\begin{align}
    h &= \frac{d}{2}-1
    &
    \nu &= \sqrt{h^2 + \mA^2 R^2}
    \ .
\end{align}
We present the self energies $\Sigma(p)$ of the transverse part of the gauge field, $A^\mu_\textnormal{T}$, for different values of $\alpha$.

\paragraph{Case: \texorpdfstring{$\alpha\gg \frac{d}{2}$}{Case: Alpha >> d/2}}
\begin{align}
    \Sigma(p)
    &= \sqrt{-p^2} 
    \frac{ 
        K_{h-1}
        (\sqrt{-p^2\zUV^2}) 
        }{
        K_h
        (\sqrt{-p^2\zUV^2}) 
        }  
    \approx 
    \frac{2}{\zUV}
    \frac{ 
        \Gamma(1-h)
        }{
        \Gamma(h)
        }
        \left( 
            \frac{ -p^2 \zUV^2 }{4}
        \right)^{ h }
         \ ,
\end{align}
where the approximation holds for $|p\zUV| \ll 1$ and $d>2$.

\paragraph{Case: \texorpdfstring{$\alpha = \frac{d}{2}$}{Case: Alpha = d/2}}
\begin{align}
    \Sigma(p)
    &= 
    - \frac{ d - 2(1 + \nu) }{ 2 \zUV } 
    + \sqrt{ -p^2 } 
        \frac{ K_{\nu-1} ( \sqrt{-p^2 \zUV^2} ) }{ 
            K_{\nu}(\sqrt{-p^2 \zUV^2} ) }  
    \\
    & \approx 
     -  \frac{p^2 \zUV }{ 2 ( \nu - 1 )}
      -
      \frac{ d - 2 (1 + \nu) }{ 2 \zUV } 
      + 
      \frac{2}{\zUV}
      \frac{\Gamma(1-\nu)}{\Gamma(\nu)}
      \left( \frac{-p^2 \zUV^2}{4} \right)^{\nu} 
      \ ,
      \label{eq:Sigma:alpha:d:2}
\end{align}
where the approximation holds for $|p\zUV| \ll 1$. The $\alpha = d/2+\varepsilon$ case  is detailed in Section \ref{eq:CFTapprox}. 

\paragraph{\texorpdfstring{Case: $\alpha = \frac{d}{2}-1$}{Case: alpha = d/2}}
\begin{align}
    \Sigma(p)
    & = 
    \sqrt{ \mA ^2 - p^2 } 
    \frac{
            K_{h-1}(\sqrt{\mA ^2-p^2} z_{\rm UV} )
        }{
            K_{h}(\sqrt{\mA ^2-p^2} z_{\rm UV} )
        } 
     \approx 
     \frac{2}{\zUV}
     \frac{\Gamma(1-h)}{\Gamma(h)}
     \left[
        \frac{(\mA ^2-p^2)z_{\rm UV}^2}{4}
    \right]^{h} 
      \label{eq:gappedselfenergyapprox:repeat}
      \ ,
\end{align}
where the approximation holds for $\left|(\mA ^2-p^2)\right| \zUV^2 \ll 1$ and  $d>2$. The approximation does not hold for integer values of $d/2$; in that case one must use the asymptotic forms of the Bessel functions of integer order.

\paragraph{\texorpdfstring{Case: $0 \leq \alpha < \frac{d}{2}-1$}{Case: alpha < d/2 - 1}} 

For $|p|\gg \mA $, the integral for the mode spacing \eqref{eq:spacing} 
is dominated by the region  near the turning point. We obtain 
\begin{align}
    \int^{z_\textnormal{c}(p)}_{\zUV} dz \frac{1}{ \sqrt{V(z)-p^2} }  
    & \approx 
    \frac{ 
        \sqrt{\pi} \Gamma( 1 + \ella )
        }{
        \Gamma( \frac{1}{2} + \ella ) 
        }
    \frac{R}{p} \left( \frac{p^2}{\mA ^2} \right)^{\!\ella}
    &
    \ella \equiv \frac{1}{d - 2 \alpha - 2}
    \,.
\end{align} 
The mode spacing behaves as $\Delta m_n \propto R^{-1}(\mA^2/m_n^2)^{\ella}$. 
We solve this to obtain the spectrum for $m_n\gg \mA $,
\begin{align}%
    m_n 
    &\approx 
        \mA
        \left[
        \frac{\sqrt{\pi} }{1+2\ella}
        \frac{  
                \Gamma( \frac{1}{2} + \ella ) 
            }{
                \Gamma(1 + \ella ) 
            }
        \left(\frac{n}{\mA  R}\right)
    \right]^{\frac{1}{1+2\ella}}
    \ .
    \label{eq:mnWKB}
\end{align}
For $d=4$, $\alpha=0$, the spectrum behaves as $m^2_n = 4 n \frac{\mA }{R}$ and thus exhibits Regge behavior. 

\paragraph{\texorpdfstring{Case: $\alpha =0$}{Case: alpha = 0}}
For $\mA  \zUV \ll 1$, the holographic self-energy of $A_\mu$  is 
\begin{align}%
    \Sigma(p)
    &= 
    \mA  
    +\frac{1}{2} \zUV p^2 
    \left[
        \gamma_\textnormal{E} 
        + \log(\mA   R)
        +H\!\left( - \frac{ p^2 R }{ 4 \mA }\right) 
    \right]
\end{align}
where $\gamma_\textnormal{E}$ is the Euler--Mascheroni constant and $H(x)$ is the analytically continued harmonic number.

\bibliographystyle{utcaps} 
\bibliography{WarpedDarkPhoton.bib}

\end{document}

%% file: FlipPreamble.tex
\usepackage{amsmath}
\usepackage{amssymb}
\usepackage{amsfonts}
\usepackage{graphicx}
\usepackage[utf8]{inputenc}     
\usepackage{aas_macros}				  
\usepackage{bm}                 
\usepackage{amsthm}
\usepackage{microtype}

\usepackage{slashed}				
\usepackage{mathrsfs}				
\usepackage{bbm}					
\usepackage{cancel}					

\usepackage[dvipsnames]{xcolor}
\usepackage[hang,flushmargin]{footmisc} 

\usepackage{fancyhdr}		
\usepackage{lipsum}			
\usepackage{mdframed}   
\usepackage{subcaption}	
\usepackage{cite}			  
\usepackage{xspace}			

\usepackage{booktabs}		

\usepackage[font=small]{caption} 	
\usepackage{float}         			  

\usepackage[margin=2.5cm]{geometry} 
\graphicspath{{figures/}}			      
\numberwithin{equation}{section}    

\usepackage{multicol}
\usepackage{etoolbox}
\usepackage{relsize}
\patchcmd{\thebibliography}
  {\list}
  {\begin{multicols}{2}\smaller\list}
  {}
  {}
\appto{\endthebibliography}{\end{multicols}}

\let\oldenumerate\enumerate
\renewcommand{\enumerate}{
  \oldenumerate
  \setlength{\itemsep}{4pt}
  \setlength{\parskip}{0pt}
  \setlength{\parsep}{0pt}
}

\let\olditemize\itemize
\renewcommand{\itemize}{
  \olditemize
  \setlength{\itemsep}{4pt}
  \setlength{\parskip}{0pt}
  \setlength{\parsep}{0pt}
}

\usepackage{lineno}

\usepackage{scalefnt} 
\newcommand\acro[1]{{\small{#1}}} 

\renewcommand{\tilde}{\widetilde}   
\renewcommand{\text}{\textnormal}	

\newcommand{\email}[1]{\href{mailto:#1}{#1}}
\newenvironment{institutions}[1][2em]{\begin{list}{}{\setlength\leftmargin{#1}\setlength\rightmargin{#1}}\item[]}{\end{list}}

\fancypagestyle{firststyle}{
	\rhead{\footnotesize%
	\texttt{\FlipTR}%
	}}

\usepackage{etoc}

%% file: FlipAdditionalHeader.tex
\usepackage{mathtools} 
\usepackage{multirow}

\newcommand{\UU}{\textnormal{\acro{U(1)}}\xspace}
\newcommand{\UV}{\acro{UV}\xspace}
\newcommand{\IR}{\acro{IR}\xspace}
\newcommand{\FiD}{\acro{5D}\xspace}
\newcommand{\FoD}{\acro{4D}\xspace}
\newcommand{\AdS}{\acro{AdS}\xspace}

\newcommand{\AdSdpp}{\acro{AdS}$_{d+1}$\xspace}
\newcommand{\CFT}{\acro{CFT}\xspace}
\newcommand{\AdSCFT}{\acro{AdS}/\acro{CFT}\xspace}
\newcommand{\WKB}{\acro{WKB}\xspace}
\newcommand{\vev}{vev\xspace}
\newcommand{\EFT}{\acro{EFT}\xspace}

\newcommand{\Phib}{\varphi}
\newcommand{\Abmu}{B^\mu}
\newcommand{\SD}{S_\textnormal{U(1)}}
\newcommand{\auv}{b_\textnormal{UV}}
\newcommand{\cuv}{c_\textnormal{UV}}
\newcommand{\vuv}{v_\textnormal{UV}}

\newcommand{\e}{e}	
\newcommand{\D}[1]{\ensuremath{\operatorname{d}\!{#1}}}
\newcommand{\Ddppx}{\operatorname{d}^{d+1}\!x}

\newcommand{\Ddx}{\operatorname{d}^{d}\!x}
\newcommand{\Ddp}{\frac{\operatorname{d}^d\!p}{(2\pi)^d}}

\newcommand{\Dz}{\operatorname{d}\!z}

\newcommand{\zUV}{z_\textnormal{UV}\xspace}
\newcommand{\zIR}{z_\textnormal{IR}\xspace}

\newcommand{\y}{y\xspace}

\newcommand{\DY}{\operatorname{d}\!\y}
\newcommand{\yt}{\y_\textnormal{t}\xspace}
\newcommand{\zt}{z_\textnormal{t}\xspace}

\newcommand{\nut}{\nu_\textnormal{t}\xspace}

\newcommand{\PhiUV}{\Phi_\textnormal{UV}\xspace}

\newcommand{\mA}{m_{\!A}}
\newcommand{\dd}{h}

\newcommand{\At}{A_\textnormal{t}\xspace}
\newcommand{\Agg}{A_{>}\xspace}
\newcommand{\All}{A_{<}\xspace}

\newcommand{\tHF}{`t~Hooft--Feynman\xspace}
\newcommand{\gauge}{\Omega\xspace}
\newcommand{\Sigmalow}{\Sigma_\textnormal{low}\xspace}
\newcommand{\p}{p} 
\newcommand{\ella}{\ell_\alpha} 

\newcommand{\g}{g}

\newcommand{\be}{\begin{equation}}
\newcommand{\ee}{\end{equation}}
\newcommand{\M}{\mathcal{M}}

\makeatletter
\newcommand{\oset}[3][0ex]{%
  \mathrel{\mathop{#3}\limits^{
    \vbox to#1{\kern-2\ex@
    \hbox{$\scriptstyle#2$}\vss}}}}
\makeatother

\usepackage{mdframed}

\global\mdfdefinestyle{flipbox}{%
	linecolor=green!50!black,linewidth=1pt,%
	leftmargin=0cm, rightmargin=0cm
}

\usepackage{tikz}
\usetikzlibrary{shapes.geometric, arrows}
\usetikzlibrary{positioning}

\tikzstyle{mynode} = [rectangle, rounded corners, 
minimum width=3cm, 
minimum height=.7cm,
text centered, 
font=\footnotesize \sffamily,
draw=black 
]

\tikzstyle{arrow} = [thick,->,>=stealth]

%% file: FlipPreambleEnd.tex
\usepackage[
	colorlinks=true,
	citecolor=green!50!black,
	linkcolor=NavyBlue!75!black,
	urlcolor=green!50!black,
	hypertexnames=false]{hyperref}